\documentclass[11pt]{article}
\usepackage{axodraw}
\input pix.sty

\def\caseX{\TopfVT(\Ahh,\Ahh,\Ahh,\Lsc,\Lsc,\Lsc)}
\def\caseIX{\TopfVT(\Ahh,\Asc,\Ahh,\Lhh,\Lhh,\Lsc)}
\def\caseVIII{\TopfVT(\Ahh,\Asc,\Asc,\Lhh,\Lsc,\Lsc)}
\def\caseVII{\TopfVT(\Asc,\Asc,\Asc,\Lsc,\Lsc,\Lsc)}

\newcommand{\Vppo}{\TopoVR(\Asc)}
\newcommand{\VppoII}{\(\!\!\TopoVR(\Asc)\!\!\)^2}
\newcommand{\VppoIII}{\(\!\!\TopoVR(\Asc)\!\!\)^3}
\newcommand{\VppoIV}{\(\!\!\TopoVR(\Asc)\!\!\)^4}
\newcommand{\VpptII}{\ToprVV(\Asc,\Asc,\Lsc,\Lhh,\Lhh)}
\newcommand{\VpptIII}{\ToprVV(\Asc,\Asc,\Lhh,\Lsc,\Lsc)}
\newcommand{\VppfIV}{\caseVII}
\newcommand{\VppfV}{\TopfVBB(\Asc,\Asc,\Asc,\Asc,\Lhh)} 
\newcommand{\VppfVI}{\caseVIII}
\newcommand{\VppfVII}{\TopfVB(\Asc,\Asc,\Asc,\Ahh,\Ahh,\Asc,\Lsc)} 
\newcommand{\VppfVIII}{\TopfVBB(\Asc,\Ahh,\Asc,\Ahh,\Lhh)} 
\newcommand{\VppfIX}{\caseX}
\newcommand{\VppfX}{\caseIX}
\newcommand{\VppfXI}{\TopfVB(\Ahh,\Asc,\Asc,\Ahh,\Asc,\Ahh,\Lsc)} 
\newcommand{\XXXX}[1]{\gamma_{#1}}
\newcommand{\us}[1]{\frac{1}{d-x}}
\newcommand{\lambdaBar}{\bar\lambda}
\newcommand{\idAt}{\frac{1}{d_A + 2}}
\newcommand{\mm}{m^2}
\newcommand{\bmu}{\bar\mu}

\renewcommand{\loop}{\ell}
\newcommand{\mum}{\frac{\bar\mu}{2m(\bmu)}}
\def\OEE{{\cal O}(\epsilon)}

\def\Ada(#1,#2)(#3,#4,#5){\DashCArc(#1,#2)(#3,#4,#5){3}}
\def\Lda(#1,#2)(#3,#4){\DashLine(#1,#2)(#3,#4){3}}
\def\intAHa{\TopoVR(\Asc)}

\def\intAHc{\ToprVB(\Asc,\Asc,\Asc,\Asc)}

\def\intYMa{\TopoVR(\Ada)}
\def\intYMb{\ToptVS(\Ada,\Ada,\Lda)}
\def\intYMc{\ToprVB(\Ada,\Ada,\Ada,\Ada)}
\def\intYMd{\ToprVV(\Ada,\Ada,\Lda,\Lda,\Lda)}
\def\intYMe{\ToprVM(\Ada,\Ada,\Ada,\Lda,\Lda,\Lda)}
\def\inta{\ToptVS(\Asc,\Asc,\Lda)}
\def\intb{\ToprVB(\Asc,\Asc,\Ada,\Ada)}
\def\ToprVBblob(#1,#2,#3,#4){\picb{#1(30,15)(15,-120,120)%
 #2(30,15)(15,120,240) #3(15,15)(15,60,300) #4(15,15)(15,-60,60)%
 \GCirc(45,15){2}{0}}}
\def\intbb{\ToprVBblob(\Asc,\Asc,\Ada,\Ada)}
\def\intc{\ToprVV(\Asc,\Asc,\Lsc,\Lda,\Lda)}
\def\intd{\ToprVV(\Asc,\Ada,\Lda,\Lsc,\Lda)}
\def\inte{\ToprVV(\Asc,\Asc,\Lda,\Lsc,\Lsc)}
\def\intf{\ToprVM(\Asc,\Asc,\Asc,\Lda,\Lda,\Lda)}
\def\intg{\ToprVM(\Asc,\Ada,\Asc,\Lsc,\Lsc,\Lda)}

\def\intB{\fr1{m^2} \(\!\! \intAHa \!\!\)^2 \times \(\!\! \intYMa \!\!\)}
\def\intC{\fr1{m^2} \(\!\! \intAHa \!\!\) \times \(\!\! \intYMa \!\!\)^2}
\def\intD{\(\!\! \intAHa \!\!\) \times \(\!\! \intYMb \!\!\)}

\def\intG{\(\!\! \inta \!\!\) \times \(\!\! \intYMa \!\!\)}
\def\intH{ \intb}
\def\intI{m^2 \intbb}

\def\intK{m^2 \intd}

\makeatletter \@addtoreset{equation}{section} \makeatother
\renewcommand{\theequation}{\arabic{section}.\arabic{equation}}
\makeatletter
\renewcommand\section{\@startsection {section}{1}{\z@}%
                                   {-5.5ex \@plus -1ex \@minus -.2ex}
                                   {2.3ex \@plus.2ex}%
                                   {\normalfont\large\bfseries}}
\renewcommand\subsection{\@startsection{subsection}{2}{\z@}%
                                     {-3.25ex\@plus -1ex \@minus -.2ex}%
                                     {1.5ex \@plus .2ex}%
                                     {\normalfont\normalsize\bfseries}}
\renewcommand\thesection {\@arabic\c@section}
\renewcommand\thesubsection   {\thesection.\@arabic\c@subsection}
\renewcommand{\@seccntformat}[1]{%
\csname the#1\endcsname.\hspace{1.0em}}
\makeatother

\begin{document}

\begin{titlepage}
\begin{flushright}
CERN-TH/2003-076\\
HIP-2003-16/TH\\
MIT-CTP 3354\\
hep-ph/0304048\\
\end{flushright}
\begin{centering}
\vfill
 
{\Large{\bf Four-Loop Vacuum Energy Density of}}

\vspace{0.2cm}

{\Large{\bf the SU($N_c$) + Adjoint Higgs Theory}} 

\vspace{0.8cm}

K. Kajantie$^{\rm a,}$\footnote{keijo.kajantie@helsinki.fi}, 
M. Laine$^{\rm b,}$\footnote{mikko.laine@cern.ch},
K. Rummukainen$^{\rm a,c,}$\footnote{kari.rummukainen@helsinki.fi},  
Y. Schr\"oder$^{\rm d,}$\footnote{yorks@lns.mit.edu}

\vspace{0.8cm}

{\em $^{\rm a}$%
Department of Physics,
P.O.Box 64, FIN-00014 University of Helsinki, Finland\\}

\vspace{0.3cm}

{\em $^{\rm b}$%
Theory Division, CERN, CH-1211 Geneva 23,
Switzerland\\}

\vspace{0.3cm}

{\em $^{\rm c}$%
Helsinki Institute of Physics,
P.O.Box 64, FIN-00014 University of Helsinki, Finland\\}

\vspace{0.3cm}

{\em $^{\rm d}$%
Center for Theoretical Physics, 
MIT, Cambridge, MA 02139, USA\\}

\vspace*{0.8cm}
 
\end{centering}
 
\noindent
We compute the dimensionally regularised four-loop vacuum energy density 
of the SU($N_c$) gauge + adjoint Higgs theory, in the disordered phase. 
``Scalarisation'', or reduction to a small set of master integrals of the 
type appearing in scalar field theories, is carried out in $d$ dimensions, 
employing general partial integration identities through an algorithm
developed by Laporta, while the remaining scalar integrals are evaluated 
in  $d = 3 - 2\epsilon$ dimensions, by expanding in $\epsilon \ll 1$ and 
evaluating a number of coefficients. The results have implications for 
the thermodynamics of finite temperature QCD, allowing to determine 
perturbative contributions of orders ${\cal O}(g^6 \ln(1/g))$, 
${\cal O}(g^6)$ to the pressure, while the general methods are applicable 
also to studies of critical phenomena in QED-like statistical physics systems. 
\vfill
\noindent
 

\vspace*{1cm}
 
\noindent
April 2003

\vfill

\end{titlepage}

%
\section{Introduction}
\la{se:introduction}

The theory we study in this paper is the Euclidean
SU($N_c$) gauge + adjoint Higgs theory, defined in continuum
dimensional regularisation by the action
\ba
 S_\rmi{E} 
 & \equiv &  \int  {\rm d}^d x\, {\cal L}_\rmi{E} 
 \;, \\
 {\cal L}_\rmi{E} 
 & \equiv & \fr12 \tr F_{kl}^2 + \tr [D_k,A_0]^2 + 
 m^2\tr A_0^2 +\lambda (\tr A_0^2)^2, 
 \la{leff}
\ea
where 
$k,l=1,...,d$,
$D_k = \partial_k - i g A_k$, 
$A_k = A_k^a T^a$, $A_0 = A_0^a T^a$, $F_{kl} = (i/g) [D_k,D_l]$, 
and $T^a$ are the Hermitean generators of SU($N_c$), 
normalised as $\tr T^a T^b = \delta^{ab}/2$.  Summation over repeated 
indices is understood.
We could have taken the scalar potential also in the form
$\lambda_1(\tr A_0^2)^2 + \lambda_2\tr A_0^4$, but the two quartic
terms are independent only for $N_c\ge 4$ and thus, to avoid
further proliferation of formulae, we will set $\lambda_2 = 0$ here, 
denoting $\lambda \equiv \lambda_1$. For the moment we keep 
$d$ general, but later on we write $d=3-2\epsilon$, and expand
in $\epsilon \ll 1$.

The observable we would like to compute for the 
theory in \eq\nr{leff} is its partition function, or 
``vacuum energy density'',  
\be
 f(m^2,g^2,\lambda) \equiv - \lim_{V \to \infty} \frac{1}{V} 
 \ln \int \! {\cal D} A_k {\cal D} A_0  
 \, \exp\Bigl( -S_\rmi{E} 
 \Bigr) \;.
\la{zdef}
\ee
Here $V$ is the $d$-dimensional volume. The phase diagram of the system
described by $S_\rmi{E}$ has a ``disordered'', or symmetric phase and, 
depending on $N_c$, various kinds of symmetry broken phases~\cite{bka,su3adj}. 
Our aim is to determine the perturbative expansion for $f$ up to 4-loop 
order in the symmetric phase, expanding around $A^a_0 = A^a_k = 0$;
the 3-loop result is known already~\cite{bn,a0cond}.
The result will depend on $N_c$ through 
$d_A\equiv N_c^2-1, C_A\equiv N_c$.

The main motivation for the exercise described comes from 
finite temperature QCD. Indeed, the 
simplest physical observable there, the free energy density
or minus the pressure, has been computed perturbatively
up to resummed 3-loop level~\cite{az,zk}, but the expansion converges 
very slowly, requiring probably temperatures $T\gg$~TeV
to make any sense at all~\cite{az,zk,bn,adjoint}. Moreover, 
at the 4-loop level the expansion breaks down completely~\cite{linde,gpy}.
Multiloop computations are not useless, though: these infrared
problems can be isolated into the three-dimensional (3d)
effective field theory in \eq\nr{leff}~\cite{dr}, and studied 
non-perturbatively there with simple lattice simulations~\cite{a0cond}. 
However, to convert the results from lattice regularisation to 3d continuum 
regularisation, and from the 3d continuum theory to the original 4d physical
theory, still necessitates a number of perturbative multiloop
``matching'' computations.

The way our computation enters this setup has been described
in~\cite{gsixg}. Combining our results with those of another
paper~\cite{sun} allows one to determine, 
as explained in~\cite{gsixg}, all the logarithmic
ultraviolet and infrared divergences entering the 4-loop free energy
of QCD. This not only fixes the last perturbatively 
computable contribution to the free energy of hot QCD~\cite{gsixg},
of order $\mathcal{O}(g^6 \ln(1/g)T^4)$,  but 
is also a step towards renormalising the non-perturbative
contributions, as determined with lattice methods~\cite{a0cond,latt}. 
Some other applications of our results
are discussed in~\se\ref{se:appl}. 

%
\section{Outline of the general procedure}
\la{se:method}

The first step of the 
perturbative computation is the generation of the Feynman diagrams. 
At 4-loop level, this is no longer a completely trivial task. In order to 
make the procedure tractable, we employ an algorithm whereby the graphs 
are generated in two sets: two-particle-irreducible ``skeleton'' graphs, 
as well as various types of ``ring'' diagrams, containing all possible
self-energy insertions. The resulting sets, with the relevant
symmetry factors, were provided explicitly in~\cite{sd}. 

It actually turns out that some of the generic graphs shown 
in~\cite{sd} do not contribute in the present computation. There are
two reasons for this. First, once the Feynman rules for the interactions
of gauge bosons and adjoint scalars are taken into account, some of the
graphs vanish at the point of colour contractions. This concerns 
particularly the ``non-planar'' topologies~\cite{ma}.
Second, all vacuum graphs which do not contain at least one massive 
(adjoint scalar) line, vanish in strict dimensional regularisation. In some
cases such a vanishing may be due to an unphysical cancellation
between ultraviolet and infrared divergences, as we will recall in
\se\ref{se:ir}, but for the moment we accept the vanishing literally. 
The remaining skeleton graphs are then as shown in~\fig\ref{fig:graphs}.
For the ring diagrams, which by far outnumber the skeleton graphs,  
we find it simpler to treat the full sets as shown in~\cite{sd}, letting 
the two types of cancellations mentioned 
above come out automatically in the actual computation. 
For completeness, the ring diagrams are reproduced in~\fig\ref{fig:rings}.

\begin{figure}[t]

\begin{eqnarray*}
 &\mbox{2-loop}:& 
 \sy{}14 \ToptVS(\Asc,\Asc,\Lgl)  
 \sy{+}18 \ToptVE(\Asc,\Asc) \nn
 &\mbox{3-loop}:&
 \sy{}16 \ToprVM(\Asc,\Asc,\Asc,\Lgl,\Lgl,\Lgl)
 \sy+18\ToprVM(\Asc,\Agl,\Asc,\Lsc,\Lsc,\Lgl) 
 \sy+12\ToprVV(\Asc,\Asc,\Lsc,\Lgl,\Lgl)  
 \sy+18\ToprVB(\Asc,\Asc,\Agl,\Agl) 
 \sy{+}1{48}\ToprVB(\Asc,\Asc,\Asc,\Asc)
 \nn
 &\mbox{4-loop}:&
 \sy-13\TopfVH(\Asc,\Agh,\Lgl,\Lgl,\Lsc,\Lsc,\Lhh,\Lhh,\Lgl)
 \sy+14\TopfVH(\Asc,\Asc,\Lsc,\Lsc,\Lgl,\Lgl,\Lgl,\Lgl,\Lgl)
 \sy+14\TopfVH(\Asc,\Asc,\Lgl,\Lsc,\Lgl,\Lsc,\Lsc,\Lgl,\Lsc)
 \sy+12\TopfVH(\Asc,\Agl,\Lsc,\Lsc,\Lgl,\Lgl,\Lsc,\Lsc,\Lgl)
 \sy+16\TopfVH(\Asc,\Agl,\Lgl,\Lgl,\Lsc,\Lsc,\Lgl,\Lgl,\Lgl)
 \sy+1{12}\TopfVH(\Asc,\Asc,\Lgl,\Lgl,\Lsc,\Lsc,\Lsc,\Lsc,\Lgl) 
 \nn[1ex]&&{}
 \sy+12\TopfVW(\Asc,\Agl,\Agl,\Agl,\Lsc,\Lsc,\Lgl,\Lgl)
 \sy+12\TopfVW(\Asc,\Agl,\Agl,\Asc,\Lgl,\Lsc,\Lgl,\Lsc)
 \sy+12\TopfVW(\Agl,\Asc,\Asc,\Asc,\Lsc,\Lsc,\Lgl,\Lgl) 
 \sy+18\TopfVW(\Asc,\Asc,\Asc,\Asc,\Lgl,\Lgl,\Lgl,\Lgl) 
 \sy+14\TopfVV(\Agl,\Agl,\Asc,\Lgl,\Lsc,\Lsc,\Lgl,\Lgl)
 \sm{+1}\TopfVV(\Asc,\Asc,\Asc,\Lsc,\Lgl,\Lgl,\Lgl,\Lgl) 
 \sm{+1}\TopfVV(\Asc,\Asc,\Agl,\Lsc,\Lsc,\Lsc,\Lgl,\Lgl) 
 \nn[1ex]&&{}
 \sy+14\TopfVB(\Agl,\Agl,\Asc,\Agl,\Agl,\Asc,\Lsc)
 \sy+18\TopfVB(\Agl,\Agl,\Asc,\Asc,\Asc,\Asc,\Lgl)
 \sy+12\TopfVB(\Asc,\Agl,\Asc,\Asc,\Agl,\Agl,\Lgl)
 \sy+12\TopfVB(\Asc,\Agl,\Agl,\Asc,\Agl,\Asc,\Lsc)
 \sy+18\TopfVB(\Asc,\Asc,\Agl,\Agl,\Agl,\Agl,\Lgl) 
 \sy+12\TopfVN(\Asc,\Asc,\Lsc,\Lsc,\Lgl,\Lgl,\Lgl) 
 \nn[1ex]&&{}
 \sy+12\TopfVU(\Asc,\Asc,\Asc,\Asc,\Lgl,\Lgl,\Lgl)
 \sy+1{16}\TopfVT(\Agl,\Asc,\Agl,\Lgl,\Lgl,\Lsc)
 \sy+16\TopfVT(\Asc,\Asc,\Asc,\Lgl,\Lgl,\Lgl) 
 \sy{+}14\TopfVW(\Agl,\Agl,\Asc,\Asc,\Lsc,\Lsc,\Lsc,\Lsc) 
 \sy+14\TopfVB(\Asc,\Asc,\Asc,\Agl,\Agl,\Asc,\Lsc)
 \sy+1{16}\TopfVT(\Asc,\Agl,\Asc,\Lsc,\Lsc,\Lgl) 
 \sy{+}18\TopfVB(\Asc,\Asc,\Asc,\Asc,\Asc,\Asc,\Lgl) 
 \sy{+}1{48}\TopfVT(\Asc,\Asc,\Asc,\Lsc,\Lsc,\Lsc) 
\end{eqnarray*}

\caption[a]{\it The skeleton diagrams contributing in~\eq\nr{zdef}, 
after subtraction of 
those which obviously vanish because of colour contractions or 
specific properties of dimensional regularisation.
Solid lines represent the adjoint scalar $A_0$, wavy lines the gauge 
boson $A_i$, and dotted lines the ghosts. The complete sets of skeleton 
diagrams have been enumerated and written down in ref.~\cite{sd}, 
whose overall sign conventions we also follow.}  
\label{fig:graphs}

\end{figure}

\begin{figure}[p]

\begin{eqnarray*}
%
%
\TopoS(\Lgl) & \equiv &
\sy{}12 \TopoSB(\Lgl,\Agl,\Agl)
\sm{-1} \TopoSB(\Lgl,\Agh,\Agh)
\sy+12 \TopoST(\Lgl,\Agl)
\sy+12 \TopoSB(\Lgl,\Asc,\Asc)
\sy+12 \TopoST(\Lgl,\Asc) \;,
\\[0ex]
\TopoS(\Lgh) & \equiv &
\sm{1} \TopoSB(\Lhh,\Agl,\Agh) \;,
\\[0ex]
\TopoS(\Lsc) & \equiv & 
\sm{1} \TopoSB(\Lsc,\Asc,\Agl) 
\sy+12 \TopoST(\Lsc,\Agl)  
\plus\sy{}12 \TopoST(\Lsc,\Asc) \;,
\\[0ex]
%
%
\ToptSi(\Lgl) & \equiv & 
\sy{}12 \ToptSM(\Lgl,\Agl,\Agl,\Agl,\Agl,\Lgl)
\sm{-1} \ToptSM(\Lgl,\Agh,\Agl,\Agl,\Agh,\Lgh)
\sm{-1} \ToptSM(\Lgl,\Agl,\Agh,\Agh,\Agl,\Lagh)
\sm{-1} \ToptSM(\Lgl,\Agh,\Agh,\Agh,\Agh,\Lgl)
\sy+12 \ToptSAl(\Lgl,\Agl,\Agl,\Agl,\Agl)
\sy+12 \ToptSAr(\Lgl,\Agl,\Agl,\Agl,\Agl) \nn[0ex]&&{}
\sy+14 \ToptSE(\Lgl,\Agl,\Agl,\Agl,\Agl)
\sy+16 \ToptSS(\Lgl,\Agl,\Agl,\Lgl)
\sy+12 \ToptSM(\Lgl,\Asc,\Agl,\Agl,\Asc,\Lsc)
\sy+12 \ToptSM(\Lgl,\Agl,\Asc,\Asc,\Agl,\Lsc)
\sy+12  \ToptSM(\Lgl,\Asc,\Asc,\Asc,\Asc,\Lgl) \nn[0ex] && {}
\sm{+1} \ToptSAl(\Lgl,\Asc,\Asc,\Asc,\Agl)
\sm{+1} \ToptSAr(\Lgl,\Asc,\Asc,\Asc,\Agl)  
\sy+12 \ToptSAl({\SetColor{Black}\Lgl},\Agl,\Asc,\Agl,\Asc)
\sy+12 \ToptSAr({\SetColor{Black}\Lgl},\Asc,\Agl,\Agl,\Asc) \nn[0ex] && {}
\sy+14 \ToptSE({\SetColor{Black}\Lgl},\Asc,\Agl,\Agl,\Asc)
\sy+14 \ToptSE({\SetColor{Black}\Lgl},\Agl,\Asc,\Asc,\Agl)
\sy+12 \ToptSS(\Lgl,\Asc,\Asc,\Lgl) 
\plus\sy{}14  \ToptSE({\SetColor{Black}\Lgl},\Asc,\Asc,\Asc,\Asc) \;,
\\[0ex]
\ToptSi(\Lgh) & \equiv &
\sm{1} \ToptSM(\Lhh,\Agl,\Agl,\Agh,\Agh,\Lgl)
\sm{+1} \ToptSM(\Lhh,\Aagh,\Agl,\Agh,\Agl,\Lagh) \;,
\\[0ex]
\ToptSi(\Lsc) & \equiv &
\sm{1} \ToptSM(\Lsc,\Agl,\Agl,\Asc,\Asc,\Lgl)
\sm{+1} \ToptSM(\Lsc,\Asc,\Agl,\Asc,\Agl,\Lsc)
\sy+12 \ToptSAl({\SetColor{Black}\Lsc},\Agl,\Agl,\Asc,\Agl)
\sy+12 \ToptSAr({\SetColor{Black}\Lsc},\Agl,\Agl,\Asc,\Agl) 
\sm{+1} \ToptSAl(\Lsc,\Asc,\Asc,\Agl,\Agl)
\sm{+1} \ToptSAr(\Lsc,\Asc,\Asc,\Agl,\Agl) \nn[0ex] && {}
\sm{+1} \ToptSE(\Lsc,\Agl,\Agl,\Asc,\Asc)
\sy+12 \ToptSS(\Lsc,\Agl,\Agl,\Lsc) 
\plus\sy{}12 \ToptSAl({\SetColor{Black}\Lsc},\Agl,\Asc,\Asc,\Asc)
\sy+12 \ToptSAr({\SetColor{Black}\Lsc},\Asc,\Agl,\Asc,\Asc) 
\plus\sy{}16 \ToptSS(\Lsc,\Asc,\Asc,\Lsc) \;,
\\[0ex]
%
%
\ToptSr(\Lgl ) & \equiv &
\sm{1} \ToptSBB(\Lgl,\Agl,\Agl)
\sm{-1} \ToptSBB(\Lgl,\Agh,\Agh)
\sm{-1} \ToptSBB(\Lgl,\Aagh,\Aagh)
\sy+12 \ToptSTB(\Lgl,\Agl)
\sm{+1}\ToptSBB(\Lgl,\Asc,\Asc)
\sy+12\ToptSTB(\Lgl,\Asc) \;,
\\[0ex]
\ToptSr(\Lgh) & \equiv &
\sm{1} \ToptSBB(\Lhh,\Agl,\Agh)
\sm{+1} \ToptSBB(\Lhh,\Aagh,\Agl) \;,
\\[0ex]
\ToptSr(\Lsc)& \equiv &
\sm{1}\ToptSBB(\Lsc,\Agl,\Asc)
\sm{+1}\ToptSBB(\Lsc,\Asc,\Agl)
\sy+12\ToptSTB(\Lsc,\Agl)
\sy+12\ToptSTB(\Lsc,\Asc) \;, \\
\mbox{3-loop:}  & &
\sy{}14  \TTopoVRoo(\Agl)
\sy-12  \TTopoVRoo(\Agh)
\sy+14  \TTopoVRoo(\Asc) \;,
\\[1ex]
\mbox{4-loop:} & &
\sy{}16  \TTopoVRooo(\Agl)
\sy+12 \TTopoVRoi(\Agl)
\sy+14  \TTopoVRor(\Agl)
\sy-13  \TTopoVRooo(\Agh)
\sm{-1} \TTopoVRoi(\Agh)
\sy-12  \TTopoVRor(\Agh)
\nn[1ex]&&{}
\sy+16  \TTopoVRooo(\Asc)
\sy+12  \TTopoVRoi(\Asc) 
\sy+14  \TTopoVRor(\Asc) \;.
\end{eqnarray*}

\caption[a]{\it The ring diagrams contributing in~\eq\nr{zdef}~\cite{sd}. 
The notation is as in~\fig\ref{fig:graphs}.}  
\label{fig:rings}

\end{figure}

The Feynman rules for the vertices and propagators appearing are 
the standard ones. We employ covariant gauge fixing, with a general
gauge fixing parameter, denoted here by
\be
 \xi \equiv \xi_\rmi{here} \equiv 1 - \xi_\rmi{standard} \;,
\ee 
where $\xi_\rmi{standard}$ is the gauge fixing parameter 
of the standard covariant gauges. Therefore, Feynman gauge
corresponds here to $\xi = 0$, Landau gauge to $\xi = 1$.
We keep everywhere $\xi$ completely general, however, and verify 
explicitly that it cancels in all the results. 

The graphs having been identified and the Feynman rules specified, 
we program them~\cite{ys_radcor} in the symbolic 
manipulation package FORM~\cite{jamv}, for further treatment. 

After the colour contractions, 
the next step is to ``scalarise'' the
remaining integrals. That is, we want to remove all scalar products from
the numerators of the momentum integrations, such that only 
integrations of the type appearing in scalar field theories remain.  
This problem can be solved
by using general partial integration identities~\cite{pi}.
The full power of the identities can be conveniently made use of through
an algorithm developed by Laporta~\cite{laporta}. We discuss some aspects 
of our implementation of this algorithm, together with the results 
obtained, in~\se\ref{se:scalar}. 

After the reduction to scalar integrals, we are faced with their evaluation. 
At this point one has to specify the dimension $d$
of the spacetime, in order
to make further progress. We write $d=3 - 2\epsilon$, 
expand in $\epsilon\ll 1$, and evaluate the various scalar integrals
appearing to a certain (integral-dependent) depth in this expansion, 
such that a specified order is achieved for the overall result. 
For the new 4-loop contributions, the overall order for which
we have either analytic or numerical
expressions is ${\cal O}(1)$. The scalar
integrals needed for this
are discussed in~\se\ref{se:integrals}.

There is one remaining step to be taken before we have the final
result: the renormalisation of the parameters $m^2,g^2,\lambda$
in~\eq\nr{leff}. In other words, the results presented up to this point
were in terms of the bare parameters, and we now want to 
re-expand them in terms of the renormalised parameters. This 
step is also specific to the dimension, and turns out to be 
particularly simple for $d=3-2\e$, since only the mass parameter
gets renormalised. The conversion of the bare parameters to the
renormalised ones is discussed in~\se\ref{se:ct}, and the final
form of the results is then shown in~\se\ref{se:final}.

Having completed the straightforward computation, we discuss the
conceptual issue of infrared divergences in~\se\ref{se:ir}. We mention 
in this context also some checks of our results, based on largely 
independent computations. We end with a list of some 
applications in~\se\ref{se:appl}. 

%
\section{Scalarisation in $d$ dimensions}
\la{se:scalar}

\begin{figure}[t]
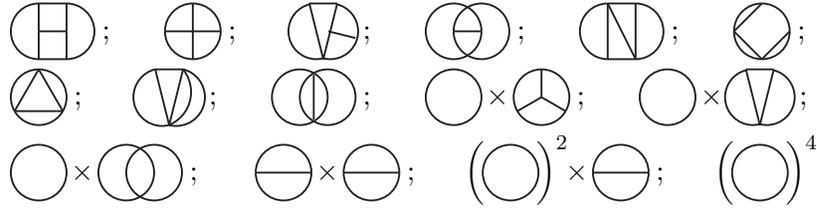


\begin{eqnarray*}
 & & 
\TopfVH(\Asc,\Asc,\Lsc,\Lsc,\Lsc,\Lsc,\Lsc,\Lsc,\Lsc) 
;\quad\;\,
\TopfVW(\Asc,\Asc,\Asc,\Asc,\Lsc,\Lsc,\Lsc,\Lsc) 
;\quad\;\, 
\TopfVV(\Asc,\Asc,\Asc,\Lsc,\Lsc,\Lsc,\Lsc,\Lsc) 
;\quad\;\,
\TopfVB(\Asc,\Asc,\Asc,\Asc,\Asc,\Asc,\Lsc) 
;\quad\;\,
\TopfVN(\Asc,\Asc,\Lsc,\Lsc,\Lsc,\Lsc,\Lsc) 
;\quad\;\, 
\TopfVU(\Asc,\Asc,\Asc,\Asc,\Lsc,\Lsc,\Lsc) 
;\nn
 & & 
\TopfVT(\Asc,\Asc,\Asc,\Lsc,\Lsc,\Lsc) 
;\quad\;\,
\TopLV(\Asc,\Asc,\Lsc,\Lsc,\Lsc,\Asc) 
;\quad\;\,
\TopfVBB(\Asc,\Asc,\Asc,\Asc,\Lsc) 
;\quad\;\,
\TopoVR(\Asc) 
\!\!\times\!\!
\ToprVM(\Asc,\Asc,\Asc,\Lsc,\Lsc,\Lsc) 
;\quad\;\,
\TopoVR(\Asc) 
\!\!\times\!\!
\ToprVV(\Asc,\Asc,\Lsc,\Lsc,\Lsc) 
;\nn 
 & &
\TopoVR(\Asc) 
\!\!\times\!\!
\ToprVB(\Asc,\Asc,\Asc,\Asc) 
;\quad\;\;
\ToptVS(\Asc,\Asc,\Lsc)
\!\!\times\!\!
\ToptVS(\Asc,\Asc,\Lsc) 
;\quad\;\;
\(\!\! \TopoVR(\Asc) \!\! \)^2 
\!\!\times\!\!
\ToptVS(\Asc,\Asc,\Lsc) 
;\quad\;\;
\(\!\! \TopoVR(\Asc) \!\! \)^4 
\end{eqnarray*}

\caption[a]{\it The 15 general types of 4-loop integrations remaining, 
in terms of momentum flow (momentum conservation is assumed at the vertices), 
after taking into account that colour 
contractions remove the non-planar topologies. Any line could contain 
a propagator to some power $n \ge 1$, and there is also an unspecified
collection of scalar products of the integration momenta in the numerator.}  
\label{fig:topologies}

\end{figure}

After inserting the Feynman rules and carrying out the colour contractions, 
there remains, at 4-loop level, a 4$d$-dimensional momentum integration to 
be carried out. The different types of integrations emerging can be 
illustrated in graphical notation in the standard way. Without specifying
the fairly complicated numerators, involving all possible kinds
of scalar products of the integration momenta, the graphs are of the
general types shown in~\fig\ref{fig:topologies}. 

There are a few simple tricks available in order
to try and simplify the scalar products appearing 
in the numerators~\cite{ys_radcor}. 
For instance, one can find relabelings of the integration variables such
that the denominators appearing in the graph remain the same, while
the scalar products in the numerators may get simplified, 
after symmetrising
between such relabelings. Some scalar products in the numerators can also
be completed into sums of squares, such that they cancel against 
the denominators. Furthermore, we can make use of various special
properties of dimensional regularisation: any closed massless 
1-loop tadpole integral vanishes; and any 1-loop massive bubble 
diagram with at most one external momentum is easily scalarised 
explicitly, in the sense of removing the loop momentum from all
the scalar products appearing in the numerators.  
However, while such
simple tricks are sufficient at, say, 2-loop level, this is 
no longer the case at 4-loop level. 

To scalarise the 4-loop integrations, we have to make full use of the 
identities provided by general partial integrations~\cite{pi}. 
To systematically employ all such
identities, we implement the algorithm presented by Laporta~\cite{laporta}
using the ``tables'' routines of FORM~\cite{jamv}. This leads to 
a complete solution of our problem. The main technical details of our
implementation were discussed in~\cite{ys_radcor}.

\begin{figure}[t]
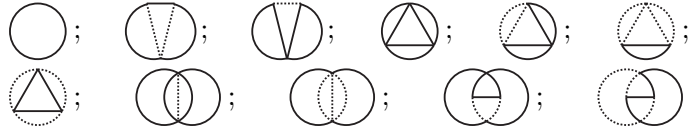


\begin{eqnarray*}
 & & 
 \Vppo
 ;\quad\,
 \VpptII
 ;\quad\,
 \VpptIII
 ;\quad\,
 \VppfIV
 ;\quad\,
 \VppfVI
 ;\quad\,
 \VppfX
 ;\nn  
 & & 
 \VppfIX
 ;\quad\;\,
 \VppfV
 ;\quad\;\,
 \VppfVIII
 ;\quad\;\,
 \VppfVII
 ;\quad\;\,
 \VppfXI
\end{eqnarray*}

\caption[a]{\it The 1-loop, 3-loop and 4-loop ``master'' 
topologies remaining after ``scalarisation''. There are no numerators left
in these graphs. A solid line is a massive propagator, $1/(p^2 + m^2)$, 
and a dotted line a massless one, $1/p^2$, where $p$ is the
Euclidean momentum flowing through the line. Note that no independent
2-loop representative appears.}  
\label{fig:masters}

\end{figure}

After the scalarisation, the master integrals remaining are those shown
in~\fig\ref{fig:masters}. This basis is, of course, not unique. As an example,
one could have chosen a different basis for the 3-loop master integrals, 
employing identities following from partial 
integrations~\cite{broadhurst_qed}, 
\ba
 \ToprVB(\Asc,\Asc,\Ahh,\Ahh) & = & 
 \frac{1}{m^2} \VppoIII
 \biggl[
 - \frac{(d-2)^2}{(d-3)(3d-8)}
 \biggr]
 +
 m^2 \VpptII
 \biggl[
 \frac{4(d-3)}{(3d-8)} 
 \biggr] \la{vpp2}
 \;,
 \\ 
 \ToprVB(\Asc,\Asc,\Asc,\Asc) & = & 
 \frac{1}{m^2} \VppoIII
 \biggl[
 -\frac{2(d-2)^2}{(d-3)(3d-8)}
 \biggr]
 +
 m^2 \VpptIII 
 \biggl[
 -\frac{4(d-4)}{(3d-8)}
 \biggr] \la{vpp3}
 \;,
\ea
where 
\be
 \Vppo \equiv \int \frac{{\rm d}^d p}{(2\pi)^d} \frac{1}{p^2+m^2} \;, 
\ee
and correspondingly for the higher loop integrals. 
Therefore, the 3-loop master integrals we are using, appearing
on the right-hand-sides of \eqs\nr{vpp2}, \nr{vpp3}, could be 
exchanged in favour of the 3-loop integrals on the left-hand-sides
of \eqs\nr{vpp2}, \nr{vpp3}.

To display the full result after scalarisation, 
we introduce the shorthand notations 
\be
 f(m^2,g^2,\lambda) 
 \equiv - d_A \sum_{\loop=1}^{\infty} (g^2 C_A)^{\loop-1} \hat p_\loop 
 \;, \qquad
 \lambdaBar \equiv \frac{\lambda (d_A+2)}{g^2 C_A} \;.
 \la{shorthand}
\ee
We then obtain the following expressions: 
\ba
 \hat p_1 & = &  m^2 \Vppo \biggl[-\frac{1}{d}\biggr] \;,
 \la{hatp1} \\
 \hat p_2 & = & \VppoII \biggl[     \frac{(d-1)}{4(d - 3)} 
                               -\fr14 \lambdaBar \biggr] \;,
 \la{hatp2} \\
 \hat p_3 & = & \frac{1}{\mm} \VppoIII \times 
        \nn  & & 
         \times \biggl\{
          \biggl[ - \frac{(d-2)(608 - 1064 d + 654 d^2 - 155 d^3 + 12 d^4)}
          {8 (d-6)(d-4)(2d-7)(3d-8)}\biggr] + 
     \nn &   & 
          + \lambdaBar \biggl[
          \frac{(d-2)(d-1)}{4(d-3)}\biggr] + 
     \nn &   & 
          + \lambdaBar^2 \biggl[- \frac{(d-2)}{8}  
          - \frac{(d-2)^2}{2 (d_A+2)(d-3)(3d-8)}  \biggr]\biggr\}  + 
     \nn & + & 
          \mm \VpptII 
          \biggl[ \frac{(d-2)^3(3d-11)}{(d-4)(2d-7)(3d-8)}\biggr] + 
     \nn & + & 
          \mm \VpptIII
          \biggl[ - \frac{(16 - 18d + 3 d^2)}{2 (d-6)(d-4)(3d-8)}
           -  \lambdaBar^2 \frac{(d-4)}{(d_A + 2)(3d-8)}\biggr]
      \;,
 \la{hatp3} \\
 \hat p_4 & = & 
        \frac{1}{m^4}\VppoIV \times 
        \nn  & & 
         \times \biggl\{   
         \frac{(d-2)\alpha_1}
         {96(d-9)(d-7)(d-6)(d-5)(d-4)^2(d-3)^3} \times
         \nn  & & \hspace*{2cm} 
         \times \frac{1}{(d-1)(2d-9)(2d-7)(3d-13)(3d-11)
         (3d-10)(3d-8)} + 
         \nn & & +  
        \lambdaBar  
        \biggl[
        - \frac{(d-2)(2904 - 7150 d + 7097 d^2 - 3581 d^3 + 964 d^4 - 
                      131 d^5 + 7 d^6 )}{8(d-6)(d-4)(d-3)^2(2d-7)} -
        \nn & & \hspace*{2cm}
        - \frac{5(d-5)(d-2)^3}{4(d_A + 2)(d-4)^2(d-3)(3d-11)}  
        \biggr] +
         \nn & & +  
         \lambdaBar^2 
         \biggl[
         \frac{(d-2)(d-1)(2d-5)}{8(d-3)}+ 
         \frac{(d-2)^2(-32 + 56 d - 25 d^2 + 3 d^3)}
         {4 (d_A + 2)(d-4)(d-3)^2(3d-8)} 
         \biggr] +
         \nn & & +  
         \lambdaBar^3
         \biggl[
         -\frac{(d-2)(2d-5)}{24} 
         -\frac{(d-2)^2}{4 (d_A+2)(d-3)}
         \biggr] \biggr\} +
         \nn & + &  
         \(\!\! \Vppo \!\!\) \times \(\!\! \VpptII \!\!\) \times
        \nn & & 
        \times \biggl\{
         \biggl[
         -\frac{(d-2)^2 \alpha_2 }
         {24(d-5)(d-4)^2(d-3)(d-1)(2d-9)(2d-7)(3d-11)(3d-10)(3d-8)} 
         \biggr] \! + \! 
         \nn & & +  
         \lambdaBar
         \biggl[ 
         \frac{(d-2)^3(3d-11)}{2(d-4)(2d-7)} 
         \biggr] \biggr\} + 
         \nn & + &  
         \(\!\! \Vppo \!\!\) \times \(\!\! \VpptIII \!\!\) \times
         \nn & & 
         \times \biggl\{ 
         \frac{\alpha_3}
         {24(d-9)(d-7)(d-6)(d-5)(d-4)(d-3)} \times 
        \nn  & & \hspace*{2cm}
          \times \frac{1}{(d-1)(2d-9)(3d-13)(3d-11)(3d-8)} + 
         \nn & & +  
        \lambdaBar
         \biggl[ 
         -\frac{16-18d+3 d^2}{4(d-6)(d-4)}
         -\frac{5(d-6)(d-2)}{(d_A+2)(d-4)(3d-11)}
         \biggr] + 
         \nn & & +  
         \lambdaBar^2
         \biggl[
         \frac{-32 + 56 d - 25 d^2 + 3 d^3}
         {2 (d_A+2)(d-3)(3d-8)}
         \biggr] + 
         \nn & & +  
         \lambdaBar^3
         \biggl[ 
         -\frac{(d-4)}{2(d_A + 2)}
         \biggr] \biggr\} + 
         \nn & + &   
         \VppfIV \times 
         \nn & & 
         \times \biggl\{ 
         \frac{\alpha_4}
         {144(d-9)(d-7)(d-5)(d-4)^2(d-1)(3d-13)(3d-11)} + 
         \nn & & +  
         \lambdaBar
         \biggl[
         -\frac{5(96 - 64 d + 7 d^2 + d^3)}
         {12(d_A+2)(d-4)^2(3d-11)}
         \biggr] + 
         \nn & & +  
         \lambdaBar^2
         \biggl[
         \frac{d}{6(d_A+2)(d-4)}
         \biggr] + 
         \nn & & +  
         \lambdaBar^3
         \biggl[
         -\frac{(d_A+8)}{6 (d_A+2)^2} 
         \biggr]
         \biggr\} + 
         \nn & + &  
         \VppfVI \times 
         \nn & & 
         \times \biggl\{
         \biggl[
         -\frac{8136 - 18176 d + 14438 d^2 - 5370 d^3 + 984 d^4 - 
         81 d^5 + 2 d^6}
         {16(d-5)(d-4)^2(d-1)(2d-9)}
         \biggr] + 
         \nn & & +  
         \lambdaBar
         \biggl[ 
         -\frac{5(20-10 d+d^2)}
         {4(d_A+2)(d-4)}
         \biggr] 
         \biggr\} + 
         \nn & + &  
         \VppfX
         \biggl[
         -\frac{(d-2)(-2656 + 5672 d - 4072 d^2 + 
         1302 d^3 - 186 d^4 + 9 d^5)}
         {16(d-4)(d-1)(3d-11)(3d-10)} 
         \biggr] + 
         \nn & + & 
         \VppfIX
         \biggl[
         -\frac{4(d-2)(-9482 + 13225 d - 7306 d^2 + 1992 d^3 - 267 d^4 + 
         14 d^5)} 
         {9(d-4)^2(2d-7)(3d-11)(3d-10)}
         \biggr] + 
         \nn & + &  
         \frac{1}{m^2} \VppfV \times 
         \nn & & 
         \times \biggl\{
         \biggl[
         -\frac{(d-2)(2d-5)\alpha_5}
         {24(d-4)^2(d-3)(d-1)(2d-9)(2d-7)} 
         \biggr] + 
         \nn & & +  
         \lambdaBar^2
         \biggl[
         -\frac{(d-2)(2d-5)}{2(d_A+2)(d-4)(d-3)}
         \biggr]  
         \biggr\} + 
         \nn & + &  
         \frac{1}{m^2} \VppfVIII
         \biggl[ 
         \frac{(d-2)(2d-5) \alpha_6}
         {3(d-4)^3(d-3)^2(d-1)(2d-7)(3d-11)(3d-10)}
         \biggr] + 
         \nn & + &  
         m^2 \VppfVII 
         \biggl\{ 
         \biggl[
         -\frac{3(11-7d+d^2)}{(d-4)(2d-9)} 
         \biggr] + 
         \lambdaBar
         \biggl[
         -\frac{10(d-3)}{(d_A+2)(d-4)} 
         \biggr]
         \biggr\} + 
         \nn & + &  
         m^2 \VppfXI 
         \biggl[ 
         \frac{2(d-3)(d-2)}{(d-4)(2d-7)}
         \biggr]
     \;, 
     \la{hatp4}
\ea
where
\ba
 \alpha_1 \!\! & = & \!\! 
  \! - \! 121583669760 
  \! + \! 2691971008704 d
  \! - \! 13463496742176 d^2
  \! + \! 33122892972480 d^3 
  \! - \!
  \nn & & \!\!
  \! - \! 50028680189824 d^4
  \! + \! 51445267135192 d^5
  \! - \! 38155599595406 d^6
  \! + \! 21131958532365 d^7
  \! - \!
  \nn & & \!\!
  \! - \! 8925676618775 d^8
  \! + \! 2909006141441 d^9
  \! - \! 734705333783 d^{10}
  \! + \! 143430052519 d^{11}
  \! - \!
  \nn & & \!\!
  \! - \! 21428725861 d^{12}
  \! + \! 2402935979 d^{13}
  \! - \! 195570319 d^{14}
  \! + \! 
  \nn & & \!\!
  \! + \! 10896768 d^{15}
  \! - \! 371376 d^{16}
  \! + \! 5832 d^{17} \;, \\
 \alpha_2 \!\! & = & \!\! 
  -14081760 + 11237380 d + 
  64451424 d^2 - 140115669 d^3 + 
  129957772 d^4 -
  \nn & & \!\!
  - 69456108 d^5 + 23323366 d^6 - 
  5020699 d^7 + 674926 d^8 - 51720 d^9 + 1728 d^{10} \;, \\ 
 \alpha_3 \!\! & = & \!\!
  508742208 - 1725645240 d + 2236030380 d^2 - 
  1426818168 d^3 + 
  \nn & & \!\!
  + 436152106 d^4 - 14158652 d^5 - 
  36636937 d^6 + 13713052 d^7 - 
  \nn & & \!\!
  - 2491870 d^8 + 
  254770 d^9 - 13967 d^{10} + 318 d^{11} \;, \\
 \alpha_4 \!\! & = & \!\! 
  -1266048 - 122112 d + 1785942 d^2 - 1171982 d^3 + 
  \nn & & \!\!
  + 307185 d^4 - 35512 d^5 + 1400 d^6 + 6 d^7 + d^8 \;, \\
 \alpha_5 \!\! & = & \!\! 
  5112 - 11321 d + 
  10618 d^2 - 5358 d^3 + 1489 d^4 - 
  212 d^5 + 12 d^6 \;, \\
 \alpha_6 \!\! & = & \!\!
  171232 - 492404 d + 584218 d^2 - 380046 d^3 + 
  \nn & & \!\!
  + 149811 d^4 - 36924 d^5 + 5595 d^6 - 480 d^7 + 18 d^8 \;. 
 \la{alpha6}
\ea

It is worth stressing that \eqs\nr{hatp1}--\nr{alpha6} were obtained with 
an arbitrary $\xi$, which just exactly 
cancelled once all the graphs were summed together, 
for a general $d$, and before inserting any properties of the master 
integrals. This is a consequence of the fact that the master integrals 
constitute really a linearly independent basis for the present problem. 

%
\section{Integrals in $d=3-2\epsilon$ dimensions}
\la{se:integrals}

A set of master scalar integrals having been identified, 
the next step is to compute them. As already mentioned, we do this by
writing $d = 3 - 2 \e$, expanding in $\e \ll 1$, and 
evaluating a number of coefficients in the series. 

In order to display the results, we first choose 
a convenient integration measure. To this end, 
we introduce an $\msbar$ scale parameter $\bmu$, 
by writing each integration as
\be
 \int \frac{{\rm d}^dp}{(2\pi)^d} \equiv 
 \mu^{-2\epsilon}\biggl[\bmu^{2\epsilon}
 \biggl( \frac{e^\gamma}{4\pi} \biggr)^\epsilon  
 \int \frac{{\rm d}^dp}{(2\pi)^d} \biggr] \;,
 \la{msb}
\ee
where $\mu = \bmu (e^\gamma/4\pi)^{1/2}$, 
and the expression in square brackets has integer dimensionality. 
This square bracket part of an $\loop$-loop integration is then written as 
\ba
 & & \hspace*{-0.5cm} 
 \biggl[\prod_{i = 1}^\loop \biggl\{ \bmu^{2\epsilon}
 \biggl( \frac{e^\gamma}{4\pi} \biggr)^\epsilon  
 \int \frac{{\rm d}^dp_i}{(2\pi)^d} \biggr\} \biggr] 
 g(p_1,...,p_\loop,m) =
 \nn & &  = 
 \frac{1}{(4\pi)^\loop} m^{3\loop-2k} 
 \biggl( \frac{\bmu}{2 m} \biggr)^{2 \e \loop} 
 \biggl\{ \prod_{i = 1}^{\loop}
 \biggl[4\pi \biggl( \frac{e^\gamma}{\pi} \biggr)^\e
 \int \frac{{\rm d}^{3 - 2\e}p_i}{(2\pi)^{3 - 2\e}}  \biggr] 
 g(p_1,...,p_\loop,1) \biggr\}  \;,
 \la{imeas}
\ea 
where $k$ counts the number of propagators, or lines, in the graphical
representation of the function $g$. 
{}From now on we assume that the loop integrations are computed
with the dimensionless measure in the curly brackets in~\eq\nr{imeas}, 
while the constants in front of the curly brackets, together with the 
explicit powers of $m$ as they appear in~\eqs\nr{hatp1}--\nr{hatp4},  
are to be provided in trivial prefactors (cf.\ \eq\nr{shorthand2} below).

With such conventions, the loop integrals remaining are 
functions of $\e$ only, and read:
\ba
 \Vppo & = & -1-2 \epsilon 
 - \epsilon^2 \Bigl( 4+\fr14 \pi^2 \Bigr) +\epsilon^3 \,\XXXX{1}
 + {\cal O}(\epsilon^4) \;, \la{ib} \\
 \VpptII & = & \frac{\pi^2}{12} +\epsilon \,\XXXX{2}
 + {\cal O}(\epsilon^2) \;, \la{XXXX2} \\
 \VpptIII & = & \ln 2 + \epsilon \,\XXXX{3}
 + {\cal O}(\epsilon^2) \;, \\
 \VppfIV & = & \frac{\pi^2}{32 \epsilon} + \XXXX{4}
 + {\cal O}(\epsilon) \;, \\
 \VppfVI & = & \frac{\pi^2}{32 \epsilon} +\XXXX{5}
 + {\cal O}(\epsilon) \;, \\
 \VppfX & = & \frac{\pi^2}{32 \epsilon}+\XXXX{6} 
 + {\cal O}(\epsilon) \;, \\
 \VppfIX & = & \frac{\pi^2}{32 \epsilon}+\XXXX{7}
 + {\cal O}(\epsilon) \;, \\
 \VppfV & = & \frac{7}{4\epsilon} -8 \ln 2 + 21 + \epsilon \,\XXXX{8}
 + {\cal O}(\epsilon^2) \;, \la{gamma8} \\
 \VppfVIII & = & \frac{3}{8 \epsilon}+\fr92+
 \epsilon \Bigl( \fr{75}2+\fr{11}8 \pi^2 \Bigr) + \epsilon^2 \,\XXXX{9}  
 + {\cal O}(\epsilon^3) \;, \la{gamma9} \\
 \VppfVII & = & \XXXX{10}
 + {\cal O}(\epsilon) \;, \la{gamma10} \\
 \VppfXI & = & {\cal O}(1) \;. \la{ie}
\ea
As we will see, the terms shown explicitly are needed for determining
the $1/\e$-poles in the 4-loop expression for $f$, 
the constants $\XXXX{n}$ are needed for 
determining the finite 4-loop contribution to $f$, and the higher order
terms only contribute at the level ${\cal O}(\epsilon)$.
Analytic results for $\gamma_1,...,\gamma_9$, as well as a numerical
determination of $\gamma_{10}$, are presented in~\ref{app:contints}.

It is now convenient to combine the conventions 
in~\eqs\nr{shorthand}, \nr{msb}, \nr{imeas} and write 
\be
 f(m^2,g^2,\lambda) = - d_A  
 \frac{\mu^{-2\e}}{4\pi}  
 \sum_{\loop=1}^{\infty} 
 m^{4 - \loop}  
 \biggl(\frac{\bmu}{2 m} \biggr)^{2 \e \loop}
 \biggl(\frac{\mu^{-2 \e}g^2 C_A}{4 \pi} \biggr)^{\loop - 1}
 \; \tilde p_\loop \;.
 \la{shorthand2}
\ee
Substituting \eqs\nr{ib}--\nr{ie} into \eqs\nr{hatp1}--\nr{hatp4}
and expanding in $\e$, 
the results then read, up to ${\cal O}(\e)$ corrections:
\ba
 \tilde p_1 
     & = &  + \fr13
     \la{tildep1} \;, \\
 \tilde p_2 
     & = & -\fr14 \biggl( \frac{1}{\e} + 3 + \lambdaBar \biggr) 
     \la{tildep2} \;,  \\
 \tilde p_3 
     & = & - \frac{89}{24} + \frac{11}{6} \ln 2 - \fr16 \pi^2  +
     \nn &  &  
     + \frac{\lambdaBar}{4} \biggl( \fr1{\e} + 3 \biggr) 
     + \frac{\lambdaBar^2}{4} \biggl[\fr12 - 
     \idAt \biggl( \fr1{\e} + 8 - 4 \ln 2 \biggr)  \biggr]
     \la{tildep3} \;, \\
 \tilde p_4 & = & 
       + \frac{1}{\e}   \biggl( \frac{43}{32} - \frac{491}{6144} \pi^2 \biggr) 
       + \frac{85291}{768} 
       - \frac{1259}{32} \ln 2 
       + \frac{5653}{1536} \pi^2 -
       \nn &  &
        - \fr14 \XXXX{1} 
        + \fr53 \XXXX{2}
        - \frac{19}{6} \XXXX{3} - \frac{157}{192} \XXXX{4} 
        - \frac{13}{64} (\XXXX{5} +  \XXXX{6} ) 
        - \fr49 \XXXX{7} 
        - \frac{19}{48} \XXXX{8} 
        - \fr16 \XXXX{9} 
        + \XXXX{10} +
       \nn & &  
       + \lambdaBar 
         \biggl[ 
         -\frac{1}{16 \e^2}
         -\frac{1}{8\e}   
         \biggl(  1 + \frac{5}{(d_A + 2)} \Bigl( \frac{\pi^2}{8} - 1 \Bigr)  
         \biggr) +
       \nn & &   \hspace*{0.8cm}
       +   \frac{37}{24} 
         - \frac{11}{12} \ln 2 
         +  \frac{\pi^2}{48} 
         + \idAt 
         \biggl( 
         - \fr52  
         - \frac{15}{2} \ln 2 
         + \frac{115}{192} \pi^2 
          - \fr54 (\XXXX{4} +  \XXXX{5} )  
         \biggr) 
         \biggr] +
       \nn & &  
       + \lambdaBar^2
         \biggl[
          \frac{1}{16 \e^2} \idAt 
       -  \frac{1}{8 \e}   
           \biggl(  1 + 
              \idAt \Bigl( 
              \frac{\pi^2}{8} 
              - 5 \Bigr)
           \biggr) -
       \nn & &  \hspace*{0.8cm}
               - \fr18 
               + \idAt \biggl( 
                 46 
               - \frac{51}{2} \ln 2 
               + \frac{13}{24} \pi^2 
               - 2 \XXXX{3} 
               - \fr12 \XXXX{4} 
               - \fr14 \XXXX{8} \biggr) 
         \biggr] + 
       \nn & &  
       + \lambdaBar^3 
         \biggl[
         \frac{1}{\e}   
          \biggl( 
          \fr18 \idAt 
          - \frac{\pi^2}{192} \frac{(d_A + 8)\,}{(d_A + 2)^2}  
          \biggr) - 
       \nn & &   
       \hspace*{0.8cm}
             - \frac{1}{24} 
             + \frac{1}{2(d_A+2)} 
               (1 - \ln 2 )  
             - \fr16 \frac{(d_A+ 8)\,}{(d_A + 2)^2} \,\XXXX{4}  
         \biggr]  
       \la{tildep4} \;.
\ea
It is interesting to note that while single diagrams contributing to 
$\tilde p_3$ do have $1/\e$-poles (cf.\ \ref{app:IRcutoff}), they sum
to zero in the term without $\lambdaBar$, but not in the terms
proportional to $\lambdaBar,\lambdaBar^2$. This structure is related 
to counterterm contributions from lower orders, as discussed in 
the next section. Similarly, single diagrams contributing to 
$\tilde p_4$ have both $1/\e^2$ and $1/\e$-poles, but the former
ones sum to zero in the term without any $\lambdaBar$'s.

Of course, single diagrams contain also $\xi$-dependence. In our 
computation $\xi$ cancelled at the stage of~\eqs\nr{hatp1}--\nr{alpha6}, 
but one could alternatively express single diagrams in terms of the 
same basis of master integrals, 
this time with $\xi$-dependent coefficients, and let  
the $\xi$'s sum to zero only in the end. For completeness, 
we again illustrate the general structure of such expressions 
at the 3-loop level, in~\ref{app:IRcutoff}.

%
\section{Counterterm contributions}
\la{se:ct}

The computation so far has been in terms of the bare
parameters of the Lagrangian in \eq\nr{leff}. As a final 
step the result is, however, to be converted into an expansion 
in terms of the renormalised parameters.  

The conversion is particularly simple in low dimensions
such as close to $d=3$, since then the 
theory in \eq\nr{leff} is super-renormalisable.
In fact, the only parameter requiring renormalisation is 
the mass parameter $m^2$. We write it as
\ba \label{mass_ct}
 m^2 & \equiv & m_\rmi{bare}^2  =  m^2(\bmu) + \delta m^2, 
 \la{baremm} \\
 \delta m^2  & = & 2 (d_A+2)   
 \frac{1}{(4\pi)^2}\frac{\mu^{-4\epsilon}
 }{4\epsilon}
 \Bigl( 
 -g^2\lambda C_A + \lambda^2 \Bigr).
 \la{mct}
\ea
This exact counterterm~\cite{framework,contlatt} 
guarantees that all $n$-point Green's functions 
computed with the theory are ultraviolet finite. 
Note that as far as dimensional reasons and single diagrams 
are concerned, there could also be divergences of the form $g^4/\e$, 
but they sum to zero in the counterterm appearing in~\eq\nr{mct}.

Inserting now~\eqs\nr{baremm}, \nr{mct} 
into the 1-loop and 2-loop expressions
for $f(m^2,g^2,\lambda)$, we get contributions of the same order
as the 3-loop and 4-loop vacuum graphs, respectively, from
$\delta m^2 \cdot \partial_{m^2} f(m^2(\bmu),g^2,\lambda)$.
We need to use here \eqs\nr{hatp1}, \nr{hatp2}, since 
$\mathcal{O}(\e)$-terms, not shown in~\eqs\nr{tildep1}, \nr{tildep2},
contribute as well, being multiplied by the $1/\e$ in $\delta m^2$.  
Explicitly, the terms to be added to \eqs\nr{tildep3}, \nr{tildep4},
once the prefactors in~\eq\nr{shorthand2} are expressed in terms of 
the renormalised parameter $m(\bmu)$ rather than $m$, are 
\ba
 \delta \tilde p_3 & = & 
 \biggl(\frac{\bmu}{2 m(\bmu)} \biggr)^{-4 \e}
 \biggl( \frac{1}{4\e} + \fr12 \biggr)
 \biggl( -\lambdaBar + \idAt \lambdaBar^2 \biggr)
 \;, \la{ct3} \\
 \delta \tilde p_4 & = & 
 \biggl(\frac{\bmu}{2 m(\bmu)} \biggr)^{-4 \e}
 \biggl( -\fr18 \biggr)
 \biggl(\frac{1}{\e^2} + \frac{1}{\e} (1 + \lambdaBar) + 
 \fr12 (4 + \pi^2) + 2 \lambdaBar  \biggr)
 \biggl( -\lambdaBar + \idAt \lambdaBar^2 \biggr)
 \;. \hspace*{0.5cm}
\ea
The 3-loop $1/\e$-contributions in \eq\nr{ct3} cancel against
the $1/\e$-terms in~\eq\nr{tildep3}. Indeed, genuine vacuum
divergences can only appear in $\tilde p_2, \tilde p_4$, since
such divergences must be analytic in the parameters $m^2,g^2,\lambda$
appearing in the Lagrangian, while $\tilde p_3$ comes with 
a coefficient $\sim (m^2(\bmu))^{1/2}$ (cf.~\eq\nr{shorthand2}). 
Another point to note is that $1/\e^2$-terms appear in $\delta \tilde p_4$
only with coefficients $\lambdaBar,\lambdaBar^2$, 
just as in~\eq\nr{tildep4}, although there is no complete
cancellation.

%
\section{The final result}
\la{se:final}

We can now collect together the full result for~$f(m^2,g^2,\lambda)$, 
in terms of the renormalised parameters of the theory. 
For dimensional reasons, its structure is, 
\ba
f(m^2,g^2,\lambda) 
  & = & \frac{\mu^{-2\epsilon}}{4\pi} 
        \;\Bigl[ \tilde f_{1,0} \Bigr] m^3(\bmu)  + 
  \nn & + & 
        \frac{\mu^{-4\epsilon}}{(4\pi)^2} \Bigl[ 
        \tilde f_{2,0}\, g^2 + \tilde f_{2,1}\, \lambda \Bigr] m^2(\bmu) + 
  \nn & + & 
        \frac{\mu^{-6\epsilon}}{(4\pi)^3} \Bigl[ 
        \tilde f_{3,0}\, g^4 + \tilde f_{3,1}\, g^2 \lambda 
        + \tilde f_{3,2}\, \lambda^2  \Bigr] m(\bmu)  + 
  \nn & + & 
        \frac{\mu^{-8\epsilon}}{(4\pi)^4} \Bigl[ 
        \tilde f_{4,0}\, g^6 + \tilde f_{4,1}\, g^4 \lambda +
        \tilde f_{4,2}\, g^2 \lambda^2 + \tilde f_{4,3}\, \lambda^3 \Bigr]+...
  \, , \la{vac_setup}
\ea
where $\tilde f_{\loop,i} = \tilde f_{\loop,i}(\epsilon, \bmu/m(\bmu))$ 
are dimensionless numbers, with $\loop$ indicating the loop order, 
and $i$ the number of $\lambda$'s appearing:
\ba \la{3loopMS}
 \tilde f_{1,0}  
 &=& d_A \Bigl( -\fr13 +\OEE \Bigr)  
 \;, \la{f10} \\
 \tilde f_{2,0} 
 &=& d_A C_A \Bigl( \fr1{4\e} +\ln\mum +\fr34 +\OEE \Bigr)
 \;,\\
 \tilde f_{2,1} 
 &=& d_A (d_A+2) \Bigl( \fr14 +\OEE \Bigr)
 \;,\\
 \tilde f_{3,0} 
 &=& d_A C_A^2 \Bigl( \fr{89}{24}-\fr{11}6\,\ln2 +\fr{\pi^2}6 +\OEE \Bigr) 
 \;,\\
 \tilde f_{3,1} 
 &=& d_A C_A (d_A+2) \Bigl( -\ln\mum -\fr14 +\OEE \Bigr)
 \;,\\
\tilde f_{3,2} 
 &=& 
 d_A(d_A+2) \Bigl( \ln\mum +\fr32 -\ln2 +\OEE \Bigr) + 
 \nn &+& 
 d_A(d_A+2)^2 \Bigl(-\fr18+\OEE\Bigr)
 \;,\\
 \tilde f_{4,0} &=&
 d_A C_A^3 \biggl[ 
 \(\frac{43}{32} - \frac{491}{6144}\pi^2 \)
 \( -\frac{1}{\e} -8 \ln \mum  \) -
 \nn&&{} 
       - \frac{85291}{768} 
       + \frac{1259}{32} \ln 2 
       - \frac{5653}{1536} \pi^2 
        + \fr14 \XXXX{1} 
        - \fr53 \XXXX{2}
        + \frac{19}{6} \XXXX{3} + 
 \nn &  &
        + \frac{157}{192} \XXXX{4} 
        + \frac{13}{64} (\XXXX{5} +  \XXXX{6} ) 
        + \fr49 \XXXX{7} 
        + \frac{19}{48} \XXXX{8} 
        + \fr16 \XXXX{9} 
        - \XXXX{10}
 + {\cal O}(\e) \biggr]
 \;, \la{eq40}\\
 \tilde f_{4,1} 
 &=&
 d_AC_A^2 \biggl[ \(\fr58 - \frac{5}{64} \pi^2 \)
  \( -\fr{1}{\e} - 8 \ln\mum \) + 
 \nn&&{}
 + \fr52 
 + \frac{15}{2} \ln 2
 - \frac{115}{192} \pi^2
 + \fr54 (\gamma_4 + \gamma_5)
 +{\cal O}(\e) \biggr] + 
 \nn&+& 
 d_A  C_A^2 (d_A+2) \biggl( -\fr1{16\e^2} 
 +\ln^2\mum 
 +\fr12 \ln\mum - 
 \nn&&{}
 - \frac{43}{24}
 + \frac{11}{12} \ln 2
 - \frac{1}{12} \pi^2
 + {\cal O}(\e)\biggr)
 \;,\\ 
 \tilde f_{4,2} 
 &=&
 d_AC_A(d_A+2) \biggl[ \fr1{16\e^2} -\fr{32-\pi^2}{64\e} 
 -\ln^2\mum -\biggl(\frac{36-\pi^2}{8}\biggr)\ln\mum - 
 \nn&&{}
 -\frac{183}{4} 
 + \frac{51}{2} \ln 2
 -\frac{23}{48} \pi^2
 + 2 \gamma_3
 + \fr12 \gamma_4
 + \fr14 \gamma_8
 +{\cal O}(\e)
 \biggr] + 
 \nn&+&
 d_AC_A(d_A+2)^2 \Bigl( 
 \fr12\ln\mum 
 -\fr18
 +{\cal O}(\e)
 \Bigr) 
 \;,\\
 \tilde f_{4,3} 
 &=& 
 d_A(d_A+2)(d_A+8)
 \biggl[ \biggl( \fr{\pi^2}{192} \biggr) \biggl( \frac{1}{\e} + 8 \ln\mum
 \biggr) 
 + \fr16 \gamma_4
 + {\cal O}(\e) 
 \biggr] + 
 \nn &+& 
 d_A(d_A+2)^2\Bigl(
 -\fr12\ln\mum
 -\fr14 + \fr12 \ln 2
 + {\cal O}(\e) 
 \Bigr) +  
 \nn &+&
  d_A(d_A+2)^3\Bigl(
 \fr1{24} +\OEE
 \Bigr)
 \;.
 \la{f43}
\ea
In particular, following the notation of ref.~\cite{gsixg} and writing
\be
 \tilde f_{4,0} \equiv - d_A C_A^3 \biggl[ 
 \alpha_\rmi{M} \biggl( 
 \frac{1}{\e} + 8 \ln \frac{\bmu}{ 2 m(\bmu)}
 \biggr)
  + \beta_\rmi{M}
 \biggr] 
 \;,
\ee
we read {}from \eq\nr{eq40} that
\ba
 \alpha_\rmi{M} & = &  \frac{43}{32} - \frac{491}{6144} \pi^2
 \approx 0.555017 \;, \\
 \beta_\rmi{M}  & = & 
       \frac{85291}{768} 
       - \frac{1259}{32} \ln 2 
       + \frac{5653}{1536} \pi^2 
        - \fr14 \XXXX{1} 
        + \fr53 \XXXX{2}
        - \frac{19}{6} \XXXX{3} -
      \nn &  &
        - \frac{157}{192} \XXXX{4} 
        - \frac{13}{64} (\XXXX{5} +  \XXXX{6} ) 
        - \fr49 \XXXX{7} 
        - \frac{19}{48} \XXXX{8} 
        - \fr16 \XXXX{9} 
        + \XXXX{10} 
      \nn  & = & 
       - \frac{311}{256}  
       - \frac{43}{32} \ln 2 
       - \frac{19}{6}\ln^2 \! 2 
       + \frac{77}{9216} \pi^2 
       - \frac{491}{1536} \pi^2 \ln 2 
       + \frac{1793}{512} \zeta(3)
       + \gamma_{10} \nn 
       & \approx &  -1.391512 
      \;. \la{bM}
\ea
In~\eq\nr{bM} we used values for $\gamma_1,...,\gamma_{10}$
{}from Appendix~\ref{app:combine} and
Appendix~\ref{app:gamma10}.
The coefficient 
$\alpha_\rmi{M}$ gives a contribution of order ${\cal O}(g^6\ln(1/g)T^4)$
and $\beta_\rmi{M}$ a perturbative contribution of order
${\cal O}(g^6T^4)$ to the pressure of hot QCD~\cite{gsixg}. 

The expression in~\eq\nr{vac_setup}, with the coefficients in 
\eqs\nr{f10}--\nr{f43}, contains a number of $1/\e^2$ and $1/\e$-poles.
Once our computation is embedded into some physical setting, 
such as in~\cite{gsixg}, a vacuum counterterm is automatically
generated (denoted by $p_\rmi{E}(T)$ in~\cite{gsixg}), which 
eventually cancels all the UV-poles, such that physical observables
remain finite for $\e\to 0$. The nature of the poles in 
\eqs\nr{f10}--\nr{f43} is analysed in detail in the next section. 

%
\section{Infrared insensitivity of the results}
\la{se:ir}

The result shown in the previous section contains
a number of $1/\e^2$ and $1/\e$-divergences. Since dimensional 
regularisation regulates at the same time both ultraviolet (UV) 
and infrared (IR) divergences, we may ask of what type are 
those obtained? The purpose of this section is to show that the 
divergences are of purely UV origin, and the result is thus IR 
insensitive, {\em if interpreted properly}. There are two ways of showing 
this, firstly an effective theory approach in which one understands that 
all the IR divergences are contained in the SU($N_c$) pure Yang-Mills 
theory obtained by integrating out the $A_0$-field, secondly 
a pragmatic one in which one shields away the IR divergences 
by giving the gluon and ghost fields a mass.

Conceptually the best way to analyse the IR sensitivity
is to dress the problem in an effective theory language. In the present
context, such an analysis was carried out in~\cite{dr}. The idea is that 
since the field $A_0$ has a mass scale, it can be integrated out.
The integration out is an ultraviolet procedure, thus by construction 
not sensitive to IR physics. The effective low-energy theory
that emerges is a 3d pure gauge theory. Its partition function, on the 
other hand, does contain IR divergences, 
starting at 4-loop level~\cite{linde,gpy}. 

Therefore, we expect that all results up to 3-loop level should 
be IR insensitive. At 4-loop level there is a part of the result, 
that is the diagrams which can be constructed fully inside the 
pure SU($N_c$) theory, which can be both IR and UV divergent. 
Since in dimensional regularisation, however, these graphs 
are set to zero, the non-zero result we have obtained 
should again be insensitive to any mass scales in the gluon 
and ghost propagators. 

Apart from the issue mentioned, there is also another possible
source of IR problems, namely that of 
overlapping divergences. Indeed, while
IR divergences appear for vacuum graphs at 
4-loop level only, they appear for self-energy graphs already
at the 2-loop level (see, e.g.,~\cite{dr}). However, 2-loop
self-energy insertions do appear also as subgraphs in the 4-loop
``ring diagrams'', making the divergence structure of such 
4-loop graphs ``doubly'' problematic. 
We return to this issue presently, but 
first finish the discussion of IR divergences at 
lower than 4-loop level. 

To be very explicit, let us 
introduce a fictitious mass parameter $m_\rmi{G}$ for all 
massless lines (gluons and ghosts), hence giving the function
$f(m^2,g^2,\lambda)$ a further functional dependence on the mass ratio
$x=m_\rmi{G}/m$. Let us denote by AH (``Adjoint Higgs'') graphs with
at least one $A_0$-line, and by YM (``Yang-Mills'') graphs with 
none at all. 
The general structure of the bare $f(m^2,g^2,\lambda)$ can then 
be expressed as (cf.\ \eq\nr{shorthand2})
\be
 f(m^2,g^2,\lambda)
 = \sum_{\loop = 1}^{\infty}
 \Bigl( \frac{\mu^{-2\e}}{4\pi} \Bigr)^\loop
 \Bigl( \frac{\bmu}{2 m} \Bigr)^{2\e\loop}
 (g^2)^{\loop -1} m^{4-\loop} 
 \Bigl[
 \tilde f_\loop^\rmi{AH}(x,\epsilon,\xi,\lambdaBar)
 + x^{4-\loop-2\e\loop}
 \tilde f_\loop^{\,\rmi{YM}}(\epsilon,\xi)  
 \Bigr]
 \;,
\ee
where $\tilde f_\loop^\rmi{AH}, \tilde f_\loop^{\,\rmi{YM}}$ are
dimensionless functions. While the treatment above corresponds to 
setting $x=0$ first and then computing the expansion in $\e$, we now 
keep a non-zero $x$ through the entire calculation, being interested 
in the limit of small $x$ only in the end: 
 \ba 
 \msbar: && 
 \lim_{\e\rightarrow 0} \lim_{x\rightarrow 0} 
 \tilde f_\loop^\rmi{AH} (x,\e,\xi,\lambdaBar)\;, \la{msbar} \\ 
 \mbox{IR-regulator}: &\;& 
 \lim_{x\rightarrow 0} \lim_{\e\rightarrow 0} 
 \tilde f_\loop^\rmi{AH} (x,\e,\xi,\lambdaBar)\;. \la{irreg} 
\ea 
These two limits do not in general commute for single diagrams, 
but should commute for the sum. Possible power IR divergences in 
single diagrams would show up as poles in $x$, while 
logarithmic ones correspond to $\ln x$.

The main technical differences in the IR regularised procedure
with respect to the $\msbar$ computation
are a more complicated scalarisation, in the absence of 
low-level
routines specific to the presence of massless lines, such 
as the so-called ``triangle rule'', and an enlarged set of master integrals. 
Furthermore, some additional diagrams contribute, which were set 
to zero from the outset in the $\msbar$ calculation, due to the
absence of any mass scale (in some subdiagram). 

As a roundup, it turns out that, starting at the 3-loop level,  
individual diagrams {\em do} indeed contain 
logarithmic as well as powerlike IR divergences, which then cancel in the 
sum, proving {\em a posteriori} the validity of the 
dimensionally regularised $\msbar$ calculation.
For completeness, we illustrate this issue in~\ref{app:IRcutoff}.

We now return to the 4-loop level. According to the discussion above, 
the full set of graphs can be divided into four sub-classes, having
potentially different IR properties: pure Yang--Mills graphs (YM) and 
those with at least one $A_0$-line (AH), with both sets further divided 
into skeletons (\fig\ref{fig:graphs}) 
and ring diagrams (\fig\ref{fig:rings}). 
The properties of the pure Yang--Mills
diagrams are discussed in~\cite{sun}, and we only state here that 
they contain both logarithmic UV as well as IR divergences, which
however exactly cancel in strict dimensional regularisation
(but not in regularisations which only regulate the UV, 
such as lattice regularisation). 
Here we then just discuss
the skeletons and rings containing at least one massive $A_0$-line.
For simplicity, we discuss explicitly only terms without
a quartic coupling $\lambda$.

We have computed the $1/\e$-divergence in the sum of 
such AH-skeletons with in total three different mass spectra:
\begin{itemize}
\item[1.] 
 As described above, whereby the $A_0$-lines carry the mass 
 parameter $m^2$, while the gluon and ghost lines are massless. 
\item[2.] 
 By giving an equal mass to all the fields: $A_0$, gluons, and 
 ghosts. The computation proceeds in complete analogy with the 
 one described in~\cite{sun}.
\item[3.]
 By setting all masses to zero, picking some line in the 4-loop 
 vacuum graph, integrating the massless 3-loop 2-point 
 function connected to that line 
 in $d$ dimensions\footnote{This problem has been solved 
 a long time ago via integration by parts; for a discussion as well 
 as an algorithmic implementation, see \cite{mincer}.}, and regulating 
 the remaining single integral by shielding the IR with a mass and
 regulating the UV via dimensional regularization.
\end{itemize}
All three methods give the same result for the
$1/\e$-pole in AH-skeletons, confirming 
its expected IR finiteness. Expressed as a contribution 
to $\tilde p_4$ in~\eq\nr{tildep4}, the divergence appearing
in the result reads
\be
 \delta [\tilde p_4] = 
 \frac{1}{3072\e}
 \biggl[
 3 ( 696 - 56 \xi - 6 \xi^2 - 5 \xi^3 )  
 -\frac{\pi^2}{4}
 ( 832 - 144 \xi + 81 \xi^2 - 15 \xi^3 + 3 \xi^4 )
 \biggr] 
 \;. \la{skels}
\ee

For the AH-rings, on the other hand, the third method does not 
work. This is due to the overlapping divergences mentioned above: 
a 2-loop 2-point function of gluons alone leads to 
logarithmic UV and IR 
divergences, and trying to carry out the final integration
by some recipe, gives generically an outcome $\sim 1/\e^2$, but 
with a coefficient dependent on what the recipe precisely was.
To cancel the $1/\e^2$-pole, not to mention to get the 
correct coefficient for the remaining $1/\e$-pole, is a very delicate
problem, which can only be guaranteed to have been solved by employing 
a fully systematic procedure. Our non-Abelian case is therefore 
qualitatively different from a pure scalar theory, where 
the problem of overlapping divergences does not emerge~\cite{bn_phi4}.
For a discussion of the cancellation of the analogue of 
the $1/\e^2$-pole in cutoff regularisation
in the pure SU($N_c$) theory, see~\cite{ma}.

On the contrary, the AH-rings can be systematically
computed with the 1st and 2nd types of mass spectra. Both procedures give 
the same result, confirming its IR insensitivity. Summing together
with~\eq\nr{skels}, we recover the $\xi$-independent
$1/\e$-pole on the first row in~\eq\nr{tildep4}.

In summary, we have verified explicitly that the only possible IR divergence
appearing in our computation is that of the pure SU($N_c$) gauge theory, 
contained in the YM-graphs. It is addressed
further in ref.~\cite{sun}.

%
\section{Discussion and conclusions}
\la{se:appl}

The main point of this paper has been the discussion of formal analytic 
techniques for, and actual results from, the evaluation of the 4-loop 
partition function of the 3d SU($N_c$) + adjoint Higgs theory 
using dimensional regularisation. The final 
result is shown in \eqs\nr{vac_setup}--\nr{f43}. We have also 
demonstrated that if interpreted as a matching coefficient --- that is, 
if the pure Yang-Mills graphs, without any adjoint scalar lines, 
are dropped, as is automatically the case in strict dimensional 
regularisation --- then the result is IR finite. Therefore, 
all IR divergences are contained in the pure Yang-Mills theory. 
We would now like to end by recalling that such techniques 
and results have also practical applications.

Perhaps the most important application is that our results 
provide two specific new {\em perturbative} contributions to 
the free energy of hot QCD, of orders $g^6\ln(1/g)T^4$, 
$g^6T^4$~\cite{gsixg}. Similarly, they provide also new 
perturbative contributions to quark number susceptibilities~\cite{av}.
Once the parameters of the 3d theory are expressed in terms of
the parameters of the physical finite temperature QCD via 
dimensional reduction, and once other contributions of the
same parametric magnitudes are added, 
this allows for instance to re-estimate
the convergence properties of QCD perturbation theory 
at high temperatures~\cite{gsixg,av,bir}. Our present computation also
contributes to the $\msbar$ scheme renormalisation of the simplest 3d 
gauge-invariant local condensates, obtained by partial derivatives 
of the action with respect to various parameters~\cite{framework}, 
and thus in principle helps in non-perturbative studies
of the pressure of high-temperature QCD~\cite{a0cond,latt}. 
It may also allow for refined analytic estimates such as Pad\'e 
resummations~\cite{ck} for the observable in~\eq\nr{zdef}. 

Let us mention that
there has recently been significant interest in somewhat more 
phenomenological approaches to QCD perturbation theory at high 
temperatures (for reviews see, e.g.,~\cite{rvs}). As far as 
we can tell our results are of no immediate use in such settings.

Another application is that 
our general procedure is relevant for studies of critical 
phenomena in some statistical physics systems. In this context 
one may either study directly the three-dimensional physical system, 
or carry out computations first in $d= 4 - \e$ dimensions, 
expand in $\e$, and then take the limit $\e\to 1$. For instance, 
some properties of the Ginzburg--Landau theory of 
superconductivity have been addressed in the former setup 
up to 2-loop level (see, e.g.,~\cite{pert}--\cite{u1big}), 
and in the latter setup, in the disordered phase, 
up to 3-loop level~\cite{kkb}. The integrals arising 
in the disordered phase are ``QED-like'' just as in our study, 
so that scalarisation and the sets of master integrals are essentially 
the same as the present ones~\cite{ys_radcor}. Moreover, 
in the case $d= 4 - \e$, 
the master integrals can be evaluated to a high accuracy utilising
the techniques introduced in~\cite{laporta2}, while for 
$d=3-2\e$ most master integrals have been evaluated in this paper.
Our methods could therefore help in reaching the 4-loop order.  

%
\section*{Acknowledgements}

We are indebted to J.A.M. Vermaseren for discussions as well as for his 
continuing efforts in optimising FORM, and to 
A.~Rajantie and A.~Vuorinen for discussions. 
This work was partly supported by the RTN network {\em Supersymmetry and 
the Early Universe}, EU contract no.\ HPRN-CT-2000-00152, and by the Academy
of Finland, contracts no.\ 77744 and 80170.  


\appendix
\renewcommand{\thesection}{Appendix~\Alph{section}}
\renewcommand{\thesubsection}{\Alph{section}.\arabic{subsection}}
\renewcommand{\theequation}{\Alph{section}.\arabic{equation}}


\section{Master integrals}
\la{app:contints}

We discuss in this Appendix the determination 
of the scalar master integrals of \eqs\nr{ib}--\nr{ie}.
They depend on one mass-scale $m$ only and are thus ``QED-like'' 
in the generalised sense that the number of 
massive lines at each vertex is even. 
Since the dependence on $m$ is trivial and has been absorbed 
into the coefficients, see \eq\nr{imeas}, $m=1$ in most of what follows.
One obtains particularly simple expansions in $3-2\e$ 
dimensions by using the integration measure 
$\int_p = \fr{(4\pi e^\gamma)^\e}{2\pi^2} \int\! d^{3-2\e}p$,
in accordance with \eq\nr{imeas}.

We first discuss briefly the various general techniques we have
employed for the evaluation of these integrals. 
The list of techniques includes:
partial integration relations between various scalar integrals, 
in analogy with those derived at 3-loop level in~\cite{broadhurst_qed}
(Appendix~\ref{app:PI}); graphs with only two massive lines, 
which can often be evaluated exactly
(Appendix~\ref{app:exact}); graphs with two or three vertices,
which can be evaluated to a sufficient depth in $\epsilon$ using
configuration space methods (Appendix~\ref{app:xspace});
and some remaining graphs, which were evaluated in momentum 
space (Appendix~\ref{app:explicit}). We combine the results
from the various techniques in Appendix~\ref{app:combine}, 
showing the actual expansions for the master integrals to the 
depths specified in~\eqs\nr{ib}--\nr{gamma9}. There is one finite
integral remaining which we have not been able to evaluate 
analytically, corresponding to~\eq\nr{gamma10}; its numerical value 
is determined in Appendix~\ref{app:gamma10}.

\subsection{Partial integration identities}
\la{app:PI}

Implementing systematically all identities following from partial 
integrations, as discussed in~\se\ref{se:scalar}, allows not only
to express all integrals in terms of a few scalar ones, 
which do not contain any non-trivial numerators, but produces also
a set of relations between the scalar integrals. As a simple
example, we may recall that the identity
\be
 0 = \sum_{k=1}^d
 \int_{p,q} 
 \frac{\6}{\6 p_k} \biggl[ 
 \frac{p_k - q_k}{(p^2+m^2)(q^2+m^2)(p-q)^2}
 \biggr] 
 \;, \la{pi1}
\ee
leads to the relation 
\be
 \int_{p,q} 
 \frac{1}{(p^2+m^2)(q^2+m^2)(p-q)^2}
 = \frac{1}{d-3}
 \int_{p} 
 \frac{1}{(p^2+m^2)^2}
 \int_{q}
 \frac{1}{(q^2+m^2)} \;. \la{pi2}
\ee
Taking furthermore into account that in dimensional regularisation
the two integrals on the 
right-hand-side of~\eq\nr{pi2} are related, we obtain
\be
 \ToptVS(\Asc,\Asc,\Lhh) =  
 \frac{1}{m^2} \VppoII
 \biggl[
 - \frac{(d-2)}{2 (d-3)}
 \biggr] \;.
\ee
Examples of similar relations at 3-loop
level were shown in~\eqs\nr{vpp2}, \nr{vpp3}, and a complete 3-loop
analysis can be found in~\cite{broadhurst_qed} (see also~\cite{zk}). 

At 4-loop level, there are obviously many more identities than 
at 3-loop level. Rather than showing a complete list we give here, 
as an example, one of the relations:
\ba
 && \TopLV(\Asc,\Asc,\Lsc,\Lhh,\Lsc,\Asc) \label{sixx}
  = 
  \frac{1}{m^2}
  \TopoVR(\Asc) 
  \!\!\times\!\!
  \ToprVB(\Asc,\Asc,\Asc,\Asc) 
  \biggl[ -\fr{d-2}{2(d-3)}  
  \biggr]
  + 
  \frac{1}{m^2}
  \VppfV
  \biggl[
  \fr{2d-5}{4(d-3)} 
  \biggr] 
  \;.
\ea
It turns out that this relation is convenient for the determination
of the 4-loop integral on the right-hand-side. 

\subsection{Integrals known exactly}
\la{app:exact}

A few of the integrals appearing can be evaluated exactly in $d$ dimensions.
This holds particularly for cases where only two massive propagators appear.
As an example, we show how this can be done in configuration space.
The massive propagator can be written as 
\be
 G(x;m_i) 
 \equiv \int\frac{{\rm d}^{3-2\e} p}{(2\pi)^{3-2\e}}
   \frac{e^{i p\cdot x}}{p^2 + m_i^2}
 = \frac{1}{(2\pi)^{\fr32-\e}}
   \Bigl( \frac{m_i}{x} \Bigr)^{\fr12-\e} K_{\fr12-\e}(m_i x)
 \;, \la{Gx}
\ee
where $K$ is a modified Bessel function, and $x$ denotes, depending on 
the context, either a $d$-dimensional vector or its modulus. On the other 
hand, the massless part of the graph converts in configuration space to
\be
 \int\frac{{\rm d}^{3-2\e} p}{(2\pi)^{3-2\e}}
   \frac{e^{i p\cdot x}}{p^\nu}
 = \frac{\Gamma(\fr32-\frac{\nu}{2}-\e)}
   {\Gamma(\frac{\nu}{2})}
   \frac{1}{2^\nu \pi^{\fr32-\e} x^{3-\nu-2\e}} \;.
\ee
We can then employ the identity
\be
 \int_0^\infty {\rm d} x\, x^\lambda K_\mu^2(x)
 = \frac{2^{\lambda-2}}{\Gamma(1+\lambda)}
 \Gamma\Bigl( \frac{1+\lambda+2\mu}{2} \Bigr)
 \Gamma^2 \Bigl( \frac{1+\lambda}{2} \Bigr)
 \Gamma\Bigl( \frac{1+\lambda-2\mu}{2} \Bigr)
 \;.
\ee
With this result, the following expressions are easily derived
(using the integration measure inside the curly brackets in~\eq\nr{imeas}):
\ba
  \ToprVB(\Asc,\Asc,\Ahh,\Ahh)
  &=& \(-\fr1{2\e}\) \fr{(4e^\gamma)^{3\e}\; 
  \Gamma(\fr12+\e) \Gamma(\fr12-\e) \Gamma(\fr12+3\e) \Gamma^2\(1+2\e\)}
  {(1-2\e)(1-6\e) \Gamma^3(\fr12) \Gamma\(1+4\e\)} \\
  &=& -\fr1{2\e}-4-\(26+\fr{25}{24}\pi^2\)\e 
  -\(160+\fr{25}3\pi^2-\fr{47}2\zeta(3)\)\e^2 +{\cal O}(\e^3) 
 \;, \la{i2} \\
  \TopfVT(\Ahh,\Asc,\Ahh,\Lhh,\Lhh,\Lsc)
  &=& \fr{\pi^2}{32\e} \(4e^\gamma\)^{4\e} 
  \fr{\Gamma^2(\fr12+\e)\Gamma^3(\fr12-\e)\Gamma^2(\fr12+3\e)
  \Gamma(\fr12-3\e)\Gamma\(1+4\e\)}
  {(1-2\e)\Gamma^8(\fr12)\Gamma^2(1-2\e)\Gamma\(1+6\e\)}  \\
  &=& \fr{\pi^2}{32} \lk \fr1\e + 2+4\ln2 
  +\(4+\fr{17}3\pi^2+8\ln2\(1+\ln2\)\)\e +{\cal O}(\e^2) \rk 
 \;, \hspace*{0.7cm} \la{gamma6} \\  
  \TopfVBB(\Asc,\Ahh,\Asc,\Ahh,\Lhh) 
  &=& \fr3{8\e} \(4e^\gamma\)^{4\e} 
  \fr{\Gamma^2(\fr12-\e)\Gamma^2(\fr12+3\e)\Gamma\(1+2\e\)\Gamma\(1+4\e\)}
  {(1-2\e)(1-4\e)(1-6\e)\Gamma^4(\fr12)\Gamma\(1+6\e\)} \\
  &=& \fr3{8\e} +\fr92 +\(\fr{75}2+\fr{11}8\pi^2\)\e 
  +\(270+\fr{33}2\pi^2-\fr{55}2\zeta(3)\)\e^2 +{\cal O}(\e^3)
 \;. \la{VBB}  
\ea
Obviously we also know
($\int_p$ is again
according to the curly brackets in~\eq\nr{imeas}):
\ba
 \TopoVR(\Asc)
 &=& \int_p \fr1{p^2+1} 
 = -\fr{(4e^\gamma)^\e}{1-2\e} \fr{\Gamma(\fr12+\e)}{\Gamma(\fr12)} \\
 &=& -1-2\e-\biggl(4+\fr{\pi^2}4\biggr)\e^2-
 \biggl(8+\fr{\pi^2}2-\fr73\zeta(3)\biggr)\e^3 +{\cal O}(\e^4)
 \;. \la{i1}
\ea

\subsection{Configuration space evaluations}
\la{app:xspace}

Even when configuration space does not allow for an exact 
evaluation of the integral, 
like in Appendix~\ref{app:exact}, it may allow for
the most straightforward way of obtaining a number of coefficients
in an expansion of the result in~$\epsilon$. This is the case
particularly if there are only two vertices in the graph. 

At $\loop$-loop level, the graphs in this class are of the form 
\be
\ToprVBl(\Asc,\Asc,\Asc,\Asc) \quad\quad
 = 
 \Bigl[4\pi\Bigl(\frac{e^\gamma}{\pi}\Bigr)^\e \Bigr]^\loop
 \frac{2 \pi^{\fr32-\e}}{\Gamma(\fr32-\e)}
 \int_0^\infty {\rm d} x\, x^{2-2\e} \prod_{i=1}^{\loop+1} G(x;m_i)
 \;, \la{bb}
\ee
where $G(x;m_i)$ is from~\eq\nr{Gx}. 
The idea (see, e.g.,~\cite{bn_phi4})
is to split the integration into two parts:
$\int_0^\infty{\rm d}x (...) = \int_0^r {\rm d}x (...) + 
\int_r^\infty{\rm d}x (...)$. The first part is performed
in $d=3-2\epsilon$ dimensions but by using the asymptotic small-$x$
form of $G(x;m_i)$, 
\be
 G(x;m_i) = \frac{\Gamma(\fr12-\e)}{4 \pi^{\fr32-\e}} 
 \frac{1}{x^{1-2\e}} 
 \biggl[ 
 1 - \Bigl(\frac{m_i x}{2} \Bigr)^{1-2\e} 
 \frac{\Gamma(\fr12+\e)}{\Gamma(\fr32-\e)} 
 + \Bigl(\frac{m_i x}{2} \Bigr)^2 
 \frac{\Gamma(\fr12+\e)}{\Gamma(\fr32+\e)} 
 + ... 
 \biggr]
 \;, \la{Gx_small}
\ee
while the latter part, which is finite, is performed by 
expanding first in $\e$ and then carrying out the remaining
integrals is $d=3$ dimensions. For instance, 
\be
 G(x;m_i) = \frac{e^{-m_i x}}{4\pi x}
 \biggl[ 
 1 - \e
 \biggl(
 \ln\frac{m_i^2 e^\gamma}{4\pi} + 
 m_i x \int_1^\infty {\rm d} y
 \ln(y^2-1) e^{(1-y) m_i x} 
 \biggr)
 + {\cal O}(\e^2)
 \biggr]
 \;. \la{Gx_large}
\ee
When the two parts are summed together and the limit $r\to 0$ is taken, 
the dependence on $r$ cancels, and we obtain the desired result. 

In the evaluation of such integrals, dilogarithms will in general
appear. Their properties have been summarised, e.g., in~\cite{dd}. 
For completeness, let us recall here that one can shift the argument of 
\ba
\dilog(x) 
 &=& -\int_0^x\!\!{\rm d}t\; \fr{\ln(1-t)}t 
 \;=\; \sum_{n>0}\fr{x^n}{n^2} 
\ea
from the intervals [-$\infty$...-1], [-1...0], [1/2...1] to  
the interval [0...1/2] via
\ba
 && 
 \dilog(x)=\dilog\(\fr1{1-x}\)-\ln(1-x)\ln(-x)+\fr12\ln^2(1-x)-\fr{\pi^2}6
 \;, \\
 && \dilog(x)=-\dilog\(-\fr{x}{1-x}\)-\fr12\ln^2(1-x)
 \;, \\
 && \dilog(x)=-\dilog(1-x)-\ln(1-x) \ln{x}+\fr{\pi^2}6 \;,
\ea
respectively. As follows from here, the dilogarithms 
satisfy, for $x>0$, 
\ba
 && \dilog(-x) + \dilog\(-\frac{1}{x}\) = -\fr12 \ln^2 \! x  - \frac{\pi^2}{6} 
 \;.
\ea
Special values include
\be
  \dilog(-1)=-\fr{\pi^2}{12}
  \;,\quad 
  \dilog(0)=0 
  \;,\quad
  \dilog\(\fr12\)=\fr{\pi^2}{12}-\fr12 \ln^2 \! 2 
  \;,\quad 
  \dilog(1)=\fr{\pi^2}6 
  \; .
\ee

Using these identities, 
and denoting $M=m_1+m_2+m_3$, we obtain for the 2-loop case
\ba
\ToptVSn(\Asc,\Asc,\Lsc) 
 &=&  
 \Bigl( \fr2{M} \Bigr)^{4\e}
 \biggl\{ \fr1{4\e} +\fr12 
  +\e \biggl[ 
  1 - \frac{\pi^2}{24} + 
  \sum_{i=1}^3
  \dilog
  \Bigl(1 - \frac{2 m_i}{M}\Bigr)
  \biggr]
 +{\cal O}(\e^2) \biggr\} 
 \;.
\ea
For the 3-loop case, now 
denoting $M=m_1+m_2+m_3+m_4$, we obtain
\ba
\ToprVBn(\Asc,\Asc,\Asc,\Asc) 
 &=& 
 -M \Bigl( \fr2{M} \Bigr)^{6\e}
 \biggl\{
 \fr1{4\e}+2+\fr12 \sum_{i=1}^4\fr{m_i}{M} \ln\fr{M}{2m_i} \nn
 & & 
  +\e \biggl[ 
  13 + \frac{3}{16}\pi^2 + 
  \sum_{i=1}^4 \biggl(
  \Bigl(1 - \frac{2 m_i}{M}\Bigr)
  \dilog
  \Bigl(1 - \frac{2 m_i}{M}\Bigr) \nn 
 & & \hphantom{+\e \biggl[13 + \frac{3}{16}\pi^2 + \sum_{i=1}^4 }
  + 4 \fr{m_i}{M} \ln\fr{M}{2m_i} 
  +\fr12 \fr{m_i}{M} \ln^2\fr{M}{2m_i}
  \biggr)
  \biggr]
  +{\cal O}(\e^2)\biggr\} .
\ea
In particular, if all masses are equal, 
\ba
 && \ToprVB(\Asc,\Asc,\Asc,\Asc)
  =  -\fr1\e -8+4\ln2 - 4 \e
  \Bigl( 13 + \frac{17}{48}\pi^2 - 8 \ln2 + \ln^2\! 2 \Bigr) 
  +{\cal O}(\e^2)  
 \;. \la{coeffA}
\ea
The case of two massless and 
two massive lines can be checked against~\eq\nr{i2}.

The 4-loop case has only been worked out to order ${\cal O}(1)$, 
rather than ${\cal O}(\e)$. Denoting now $M = m_1+m_2+m_3+m_4+m_5$, 
\ba
\TopfVBBnn(\Asc,\Asc,\Asc,\Asc,\Lsc) \label{bbb}
 &=& {M^2} \Bigl( \fr2{M} \Bigr)^{8\e} 
  \biggl[
  \fr1{16\e} + \fr34  
  +\sum_{i\neq j} \frac{m_im_j}{2M^2}
  \Bigl(
  \fr1{8\e} + \fr32 + \ln\frac{M}{2m_i} 
  \Bigr) +{\cal O}(\e) 
  \biggr] .
\ea
The case of three massless and 
two massive lines can be checked against~\eq\nr{VBB}.
The next order, ${\cal O}(\e)$, could
also be worked out and is indeed
needed for $\gamma_8$ in~\eq\nr{gamma8}, but we choose to use another 
way to determine it, based on~\eq\nr{sixx}. 

When there are more than two vertices in the graph, the configuration
space technique gets rapidly more complicated, due to the difficult
structure of the angular integrals. There is one graph we are 
interested in, however, whose divergent and, most incredibly, also 
the constant part~\cite{aleksi} can still be obtained analytically: 
\ba
 \TopLVnn(\Asc,\Asc,\Lsc,\Lsc,\Lsc,\Asc) & = &  
 \Bigl[4\pi\Bigl(\frac{e^\gamma}{\pi}\Bigr)^\e \Bigr]^4
 \int {\rm d}^{3-2\e} x \int {\rm d}^{3-2\e} y \;
 G(x-y;m_1) \prod_{i=2}^3 G(x;m_i) 
 \prod_{j=4}^6 G(y;m_j) 
 \;. \hspace*{1cm} 
\ea
Employing the angular integral~\cite{gkp}
\be
 \int {\rm d} \Omega_y \frac{K_{\lambda}(|x-y|)}{|x-y|^{\lambda}}
 = \frac{(2 \pi)^{\lambda+1}}{(xy)^{\lambda}}
 \Bigl[
 \theta(x-y) K_{\lambda}(x) I_{\lambda}(y)  
 + \theta(y-x) K_{\lambda}(y) I_{\lambda}(x)  
 \Bigr] 
 \;,
\ee
where $\lambda = \fr12-\e$ and
on the right-hand-side $x\equiv |x|$, $y\equiv |y|$, 
one is left with two independent radial integrations which 
can be handled as above~\cite{aleksi}, by splitting the 
integrations as $\int_0^\infty{\rm d}x (...) = \int_0^r {\rm d}x (...) + 
\int_r^\infty{\rm d}x (...)$. Denoting $M_{123} = m_1+m_2+m_3$, 
$M_{23456} = m_2+m_3+m_4+m_5+m_6$, the outcome is 
\ba
\TopLVnn(\Asc,\Asc,\Lsc,\Lsc,\Lsc,\Asc) \label{lv}
 &=& \(\frac{2}{M_{123}}\)^{8\e} \fr1{32} 
  \biggl[
  \fr1{\e^2}+\fr8\e
  +4\,\phi\(\fr{M_{123}}{M_{23456}},\fr{2m_2}{M_{123}},\fr{2m_1}{M_{123}}-1\) 
  +{\cal O}(\e) 
  \biggr], \hspace*{0.5cm}
\ea
where
\ba
 \phi(x,y,z) & = & 
 13+\fr7{12}\pi^2  -4 \ln^2\! x + 
 \nn & + & 
 2\dilog(1-y)+2\dilog(y+z)+2\dilog(-z) 
 +8\fr{1-x}{x(1+z)}\dilog(1-x) + 
 \nn & + & 
   8\(1+\fr{1-x}{x(1+z)}\) 
  \biggl(
  \dilog(-xz)+\ln\! x \ln(1+xz)-\fr{\pi^2}6
  \biggr) 
  \;.
\ea
In particular, 
\ba
  && \TopLV(\Asc,\Asc,\Lsc,\Lhh,\Lsc,\Asc) \label{six}
  = \fr1{32} \lk \fr1{\e^2}+\fr8\e 
  +4\( 13-8\ln^2 \! 2 -\fr{13}{12}\pi^2\) +{\cal O}(\e) \rk 
  \;.
\ea

\subsection{Momentum space evaluations}
\la{app:explicit}

When the graph has more than two vertices, the configuration space
method is in general no longer practical. Some of these graphs are, 
however, rather easily evaluated in momentum space. This is the case
particularly for the ``triangle'' topology, shown in~\eq\nr{triangle} 
below. 

The triangle graph consists of three consecutive 1-loop self-energy
insertions, 
\be
 \int\frac{{\rm d}^{3-2\e} q}{(2\pi)^{3-2\e}}
 \frac{1}{[q^2+m_1^2][(q+p)^2+m_2^2]} = 
 \frac{\Gamma(\fr12 + \e)}{(4\pi)^{\fr32-\e} p^{1+2\e}} B(p,m_1,m_2,\e)
 \;, \la{bub}
\ee
where $B(p,m_1,m_2,\epsilon)$ is a one-dimensional 
integral over a Feynman parameter. It has the properties
\ba
 B(0,m_1,m_2,\epsilon) & = & 0, \\
 B(p,0,0,\epsilon) & = & 
 \lim_{p\to\infty}
 B(p,m_1,m_2,\epsilon) = \frac{\Gamma^2(\fr12-\e)}{\Gamma(1-2\e)}, \\
 B(p,m_1,m_2,0) & = & 2 \arctan \frac{p}{m_1 + m_2}
 \;. \la{B_lims}
\ea
The triangle graph is then just a one-dimensional integration
over the modulus of $p$. Carrying out one partial integration 
and expanding in $\e$, one obtains~\cite{aleksi} 
\ba
\TopfVTnn(\Asc,\Asc,\Asc,\Lsc,\Lsc,\Lsc) \label{tria}
 &=& \Bigl( \frac{2}{m_1+m_2}\Bigr)^{8\e} \fr{\pi^2}{32} \lk 
  \fr1\e + 2 + 4 \ln 2 
  - \chi\biggl(\frac{m_3+m_4}{m_1+m_2},\frac{m_5+m_6}{m_1+m_2}\biggr)
 +{\cal O}(\e) \rk
 \;, \hspace*{0.5cm}
 \la{triangle}
\ea
where
\ba
 && \chi(x,y)=\fr{64}{\pi^3}\int_0^\infty\!{\rm d}p \ln p\; 
 \frac{{\rm d}}{{\rm d}p}
 \lk\atn(p) \atn\(\fr{p}x\) \atn\(\fr{p}y\) \rk \;, \\
 && \chi(1,1)=\fr{84}{\pi^2}\zeta(3) \;,\quad \chi(1,0)=\fr{56}{\pi^2}\zeta(3) 
  \;,\quad \chi(0,0)=0  
 \;. \la{fs}
\ea

\subsection{Summary of expansions for master integrals}
\la{app:combine}

Given the results of the previous sections, we can collect
together the expressions for the constants $\gamma_1,...,\gamma_9$
defined in~\eqs\nr{ib}--\nr{gamma9}. From \eq\nr{i1}, 
\ba
 && \gamma_1 = -8-\fr{\pi^2}2+\fr73\zeta(3) \;.
\ea
Combining \eq\nr{vpp2} with \eq\nr{i2}, 
\be
 \gamma_2 = 
 \fr16 \pi^2 -\fr52 \zeta(3) \;.
\ee
Combining \eq\nr{vpp3} with \eq\nr{coeffA}, 
\be
 \gamma_3 = - \fr16 \pi^2 - \ln^2\! 2 \;.
\ee
{}From \eqs\nr{triangle}, \nr{fs}
(or, for $\gamma_6$, from \eq\nr{gamma6}), 
\ba
 \gamma_4 & = & \frac{\pi^2}{32} \Bigl( 
 2+4\ln2-\fr{84}{\pi^2}\zeta(3)
 \Bigr) \;, \\
 \gamma_5 & = & \frac{\pi^2}{32} \Bigl( 
 2+4\ln2-\fr{56}{\pi^2}\zeta(3)
 \Bigr) \;, \\
 \gamma_6 & = & \frac{\pi^2}{32} \Bigl( 
 2+4\ln2
 \Bigr) \;, \\
 \gamma_7 & = & \frac{\pi^2}{32} \Bigl( 
 2+12\ln2-\fr{84}{\pi^2}\zeta(3)
 \Bigr) \;.
\ea
Combining \eqs\nr{six}, \nr{sixx}, 
\be
 \gamma_8 = 
 175 - 96 \ln 2 + 16 \ln^2\! 2 + \frac{53}{12} \pi^2 \;.
\ee
Finally, from \eq\nr{VBB}, 
\be
 \gamma_9 = 
 270+\fr{33}2\pi^2-\fr{55}2\zeta(3)
 \;.
\ee

\subsection{Numerical evaluation of $\gamma_{10}$}
\la{app:gamma10}

It can easily be verified that the integrals in 
\eqs\nr{gamma10}, \nr{ie} are both infrared and ultraviolet finite. 
They can therefore be evaluated directly in $d=3$ dimensions. 
For the present application we only need $\gamma_{10}$, defined
by~\eq\nr{gamma10}. 

There is no obvious partial integration relation whereby
$\gamma_{10}$ could be reduced to a simpler integral. Due 
to the fact that there are four vertices, it is also not 
easily treated in configuration space. The most straightforward
approach seems then to be to combine the self-energy of~\eqs\nr{bub}, 
\nr{B_lims}
with the 2-loop self-energy 
\be
 \ToptSM(\Lsc,\Asc,\Ahh,\Ahh,\Asc,\Lsc)\;,
\ee 
for which a one-dimensional integral representation 
has been given in~\cite{ar_3loop}. This leads to a simple
two-dimensional integral representation:
\be
 \gamma_{10} = \frac{2}{\pi} 
 \int_0^\infty {\rm d} p\; p \,\arctan\frac{p}{2} \cdot \Pi_2(p)
 \;,  
\ee
where~\cite{ar_3loop}
\ba
 \Pi_2(p) & = & \frac{1}{p^3} 
 \int_{x_-(p)}^1 \frac{{\rm d} x}{\sqrt{p^2 x^2 - (1-x^2)^2}}
 \biggl\{
 \frac{p}{1-x^2}
 \biggl[
 (1+x^2) \arctan\frac{p}{2} - 
 \nn & & \hspace*{1cm}
 - 2 (1-x + x^2)\arctan\frac{p}{1+x}  
 + \frac{\pi}{2} (1-x)^2 \biggr] + x \ln \Bigl[ 1 + \frac{p^2}{(1+x)^2} 
 \Bigr]
 \biggr\}
 , \hspace*{0.8cm}
 \la{Pi2}
\ea
and $x_-(p)\equiv (1 + p^2/4)^{\fr12} - p/2$. We may note that 
in~\eq\nr{Pi2} it is numerically advantageous to change the integration 
variable from $x$ to $y \equiv \sqrt{x-x_-(p)}$. The final result reads
\be
 \gamma_{10} \approx 0.171007009753(1) \;,
\ee
where the number in parentheses indicates the uncertainty in the last digit. 

%
\section{Three-loop results with and without an IR cutoff}
\la{app:IRcutoff}

\begin{figure}[t]
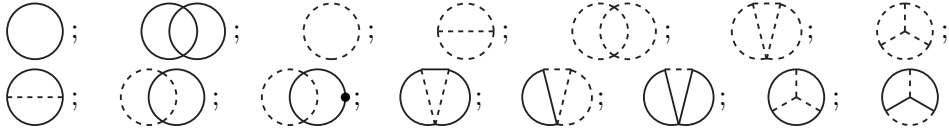


\begin{eqnarray*}
 & & 
 \intAHa 
 ;\quad\;\;\,
 \intAHc
 ;\quad\;\;\,
 \intYMa
 ;\quad\;\;\,
 \intYMb
 ;\quad\;\;\,
 \intYMc
 ;\quad\;\;\,
 \intYMd 
 ;\quad\;\;\,
 \intYMe
 ;\nn  
 & & 
 \inta 
 ;\quad
 \intb 
 ;\quad
 \intbb 
 ;\quad
 \intc 
 ;\quad
 \intd 
 ;\quad
 \inte 
 ;\quad
 \intf 
 ;\quad
 \intg
 \end{eqnarray*}

\caption[a]{\it A possible choice for 
1-loop, 2-loop and 3-loop ``master'' topologies, 
in the case that gluons and ghosts
are treated as particles with a mass $m_\rmi{G}$. There are no numerators 
left in these graphs. A solid line is a propagator of the form 
$1/(p^2 + m^2)$, and a dashed line of the form $1/(p^2+m_\rmi{G}^2)$, 
where $p$ is the Euclidean momentum flowing through the line. A line 
with a blob on it indicates a squared propagator, $1/(p^2+m^2)^2$.}  
\label{fig:mass_masters}

\end{figure}

As discussed in~\se\ref{se:ir}, starting at the 3-loop level
single graphs are considerably more infrared sensitive than the 
total sum: the limits in~\eqs\nr{msbar}, \nr{irreg} commute only
for the latter. Let us recall that there $x = m_\rmi{G}/m$, where
$m_\rmi{G}$ is a fictitious mass given to the gluons and ghosts: 
\be
 \langle A^a_k (p) A^b_l(-p) \rangle 
 \equiv 
 \frac{\delta^{ab}}{p^2 + m_\rmi{G}^2}
 \biggl( 
 \delta_{kl} - p_k p_l \frac{\xi}{p^2 + m_\rmi{G}^2}
 \biggr) \;, \quad
 \langle  c^a(p) {\bar c}^{\,b}(p)  \rangle
 \equiv 
  \frac{\delta^{ab}}{p^2 + m_\rmi{G}^2}
 \;, 
 \la{propags}
\ee
where $A^a_k, c^a, {\bar c}^{\,b}$ 
are the gluon, ghost, and anti-ghost fields, 
respectively. We illustrate the general structures appearing here with 
a few specific examples.  

In the presence of the two mass scales $m,m_\rmi{G}$, 
the set of master integrals
is more complicated than when gluons and ghosts are massless. 
The master integrals that can appear in principle are shown in 
\fig\ref{fig:mass_masters}, up to 3-loop level. 

It turns out that the skeleton diagrams are better behaved 
in the IR than the ring diagrams: power and logarithmic IR divergences 
appear only in single rings, but they cancel in their sum. 
A rather typical example, with 
both an UV pole $1/\e$ and an IR divergence $\ln(m_\rmi{G}/m)$, is
given by the gluon ring with a scalar and ghost bubble attached to it. 
Carrying out scalarisation to the master integrals 
shown in~\fig\ref{fig:mass_masters}, denoting $x = m_\rmi{G}/m$, 
and normalising as in~\eq\nr{hatp3}, 
we obtain
\ba
 \left. \sy{-}14
 \ToprVD(\Asc,\Agh,\Lgl,\Lgl,\Lgh,\Lsc) \right|_{x\neq 0} 
 & = &
  \intB \biggl[-  
  \frac{\left( d -2 \right) 
       }{8 \left(d -1 \right)  x^2} \biggr] + 
 \nn & +  & 
   \intC \biggl[  
   \frac{\left(d -2 \right) }{432 \left(d -1 \right) x^2} \biggr] \times    
     \Bigl( 54 + 54 d + 18 \xi - 18 d^2 \xi -  
 \nn & & \hphantom{
   \intC \biggl[  
   \frac{\left(d -2 \right) }{432 \left(d -1 \right)}  }
       - 14 {\xi}^2 + 14 d {\xi}^2 - d^2 {\xi}^2 + 
       d^3 {\xi}^2 \Bigr) +  
 \nn & + & 
      \intG \biggl[  
      \frac{
     \left( 20 - 8 d - 6 x^2 + 3 d x^2 \right) }{16 
     \left( d -1 \right)  x^2} \biggr] + 
 \nn & +  & 
 \intD \biggl[
 \frac{ 1 }{1296 \left( d -1\right) }
 \biggr] \times
     \Bigl( 1134 - 459 d + 54 d^2 + 108 \xi -  
 \nn & & \hphantom{
       XXXX }
       - 90 d \xi  
       - 18 d^3 \xi
       - 84 {\xi}^2 + 
       76 d {\xi}^2 + 5 d^2 {\xi}^2 + 2 d^3 {\xi}^2 + 
       d^4 {\xi}^2 \Bigr) + 
  \nn & + & 
 \intH \biggl[
     - \frac{
     \left( 32 - 12 d - 6 x^2 + 3 d x^2 \right) }{32 
     \left( d -1 \right)  x^2} \biggr] +
 \nn & +  & 
 \intI \biggl[
  - \frac{\left( -1 + x^2 \right) }
   {2 \left( d -1 \right)  x^2} \biggr] + 
 \nn & + & 
 \intK \biggl[- 
     \frac{
     \left( 44 - 8 d - 14 x^2 + 5 d x^2 \right) }{32
     \left( d -1 \right) } \biggr]
 \;.
\ea
According to~\eq\nr{irreg}, the first step is now to expand in 
$\e\ll 1$. The integrals emerging are all known~\cite{zk,ar_3loop}. 
Changing the normalisation to be according to~\eq\nr{tildep3}, we obtain 
\ba
 \left. \delta \tilde p_3 \right|_{x\neq 0}
 \!\! & = & \!\!
 -\fr1{32\e}\,(\xi-1)^2 
 +\fr{20+x^2}{128x}\biggl[ -\dilog\(\fr{3x}{2(1+x)}\)
  -\dilog\(-\fr{x}{2+x}\)-  
 \nn&&{}
  -\ln(1-\fr{x}2)\ln\fr{3x}{2(1+x)}
  -\fr12\ln^2(1+x) +\fr12\ln(1+\fr{x}2)\ln\fr{9x^2}{2(2+x)} 
  \biggr]- 
  \nn&&{} 
  -\fr{4-3x^2}{32x}\ln(1+\fr{x}2) 
  -\fr{(1+x)(3x-4)}{32x}\ln(1+x) 
  +\fr{3x}{32} + 
 \nn&&{}
  +\fr1{32}(2\xi-3)(2\xi-1)\ln\fr{3x}2  
  - \frac{837 - 954\xi + 409 \xi^2}{2592}
  +{\cal O}(\e)
 \;.
\ea
The second step is then to expand in $x \ll 1$:
\ba
 \left. \delta \tilde p_3 \right|_\rmi{\eq\nr{irreg}}
 \!\! & = & \!\!
 -\fr1{32\e} (\xi-1)^2
 + \fr{2-2\xi+\xi^2}8 \ln\fr{3x}2
 -\fr{1080-954\xi+409\xi^2}{2592}+{\cal O}(x,\e)
 . \hspace*{0.7cm} \la{ex_ir}
\ea
We observe that there is a gauge-parameter dependent UV-divergence
in the form of $1/\e$, and a gauge-parameter dependent logarithmic 
IR divergence in the form of $\ln x$. 

Proceeding according to~\eq\nr{msbar}, on the other hand, leads to
\ba
 \left. \sy{-}14
 \ToprVD(\Asc,\Agh,\Lgl,\Lgl,\Lgh,\Lsc) \right|_{x=0}
 & = & 
 \frac{1}{\mm} \VppoIII 
 \biggl[
 - \frac{(d-2)^2}{16 (d-3)(2d-7)(3d-8)} 
 \biggr] +
 \nn & + &  
 \mm \VpptII 
 \biggl[ 
 \frac{(d-3)}{4 (2d-7)(3d-8)}
 \biggr]
 \;,
\ea 
in terms of the master integrals in~\fig\ref{fig:masters}.
Expanding in~$\e\ll 1$, 
\be
 \left. \delta \tilde p_3 \right|_\rmi{\eq\nr{msbar}} = 
 \fr1{32\e} + \fr18 + \mathcal{O}(\e) 
 \;.
 \la{ex_ms}
\ee
Clearly \eqs\nr{ex_ir}, \nr{ex_ms} do not agree\footnote{%
   The first two terms in~\eq\nr{ex_ir} can be written as
   $(2-2\xi+\xi^2)[\ln(3x/2)-1/(4\e)]/8+1/(32\e)$, showing
   that the result of~\eq\nr{ex_ms} arises after a cancellation
   of IR and UV divergences in dimensional regularisation.
   }.
Summing all the graphs together, however, both procedures lead 
to the gauge-parameter independent and UV and IR finite 
$\tilde p_3$ on the first row in \eq\nr{tildep3}: in other words, 
$\xi$, $1/\e$ and $\ln{x}$ all cancel.  

Some other rings lead also to $1/x$-divergences. 
Let us show, as an example, 
\ba
  \left. \sy{}1{16}
  \ToprVD(\Asc,\Asc,\Lgl,\Lgl,\Lsc,\Lsc)
  \right|_{x\neq 0} & \Rightarrow & 
  \left. \delta \tilde p_3 \right|_\rmi{\eq\nr{irreg}} = 
  \fr1{x} \frac{24 - 12 \xi + 5 \xi^2}{64} 
  -\fr5{24} + \fr13 \ln 2 + \mathcal{O}(x,\e) \;,    
 \la{ex_ir_2}
\ea
while
\ba
  \left. \sy{}1{16}
  \ToprVD(\Asc,\Asc,\Lgl,\Lgl,\Lsc,\Lsc)
  \right|_{x=0} & \Rightarrow & 
  \left. \delta \tilde p_3 \right|_\rmi{\eq\nr{msbar}} = 
  -\fr5{24} + \fr13 \ln 2 
  + \mathcal{O}(\e) 
 \;.
 \la{ex_ms_2}
\ea
Again, 
the $1/x$-divergences of the type in~\eq\nr{ex_ir_2} cancel when gluon 
rings with all possible 1-loop scalar insertions are summed together. 

As a comparison of~\eqs\nr{ex_ir} and \nr{ex_ms},
or \eqs\nr{ex_ir_2} and \nr{ex_ms_2} shows, the computation carried
out with an IR cutoff leads in general to a more 
pronounced gauge-parameter dependence
for single graphs than the computation carried out according to~\eq\nr{msbar}, 
just because the introduction of a mass according to~\eq\nr{propags} breaks
gauge invariance. The results of~\eqs\nr{ex_ms}, \nr{ex_ms_2} 
are anomalously simple, 
however: in general there is certainly gauge-parameter dependence 
left over in single graphs also with the procedure of~\eq\nr{msbar}. 
For example, 
\ba
  \left. \sy{}18
  \ToprVD(\Asc,\Agl,\Lgl,\Lgl,\Lgl,\Lsc)
  \right|_{x=0} & \Rightarrow & 
  \left. \delta \tilde p_3 \right|_\rmi{\eq\nr{msbar}} = 
  \fr1{32 \e} (13 - 4 \xi + \xi^2 )   
  +\frac{40 - 28 \xi + 5 \xi^2}{32}
  + \mathcal{O}(\e) \;,    
\ea
and $\xi$ cancels only in the sum.



\begin{thebibliography}{99}

\bibitem{bka}
S.~Bronoff, R.~Buffa and C.P.~Korthals Altes,
{\it Phase diagram of 3D SU(3) gauge-adjoint Higgs system},
hep-ph/9809452;
%
S.~Bronoff and C.P.~Korthals Altes,
{\it Phase diagram of 3D SU(3) gauge-adjoint Higgs system
and C-violation  in hot {QCD}},
Phys.\ Lett.\ B {448} (1999) 85
[hep-ph/9811243].

\bibitem{su3adj}
A.~Rajantie,
{\it SU(5) + adjoint Higgs model at finite temperature},
Nucl.\ Phys.\ B {501} (1997) 521
[hep-ph/9702255];
%
K.~Kajantie, M.~Laine, A.~Rajantie, K.~Rummukainen and M.~Tsypin,
{\it The phase diagram of three-dimensional SU(3) + adjoint Higgs theory,}
JHEP {9811} (1998) 011 [hep-lat/9811004].

\bibitem{bn}
E. Braaten and A. Nieto,
{\it Free energy of QCD at high temperature,}
Phys.\ Rev.\ D 53 (1996) 3421 [hep-ph/9510408].

\bibitem{a0cond}
K.~Kajantie, M.~Laine, K.~Rummukainen and Y.~Schr\"oder,
{\it How to resum long-distance contributions to the QCD pressure?,}
Phys.\ Rev.\ Lett.\  {86} (2001) 10
[hep-ph/0007109].

\bibitem{az}
P.~Arnold and C.~Zhai,
{\it The three loop free energy for pure gauge QCD,}
Phys.\ Rev.\  {D 50} (1994) 7603
[hep-ph/9408276];
%
{\it The three loop free energy for high temperature QED 
and QCD with fermions,}
{\it ibid.}\  {51} (1995) 1906
[hep-ph/9410360].

\bibitem{zk}
C.~Zhai and B.~Kastening,
{\it The free energy of hot gauge theories with fermions through $g^5$,}
Phys.\ Rev.\  {D 52} (1995) 7232 [hep-ph/9507380].

\bibitem{adjoint}
K.~Kajantie, M.~Laine, K.~Rummukainen and M.~Shaposhnikov,
{\it 3d SU(N) + adjoint Higgs theory and finite-temperature QCD,}
Nucl.\ Phys.\ B {503} (1997) 357
[hep-ph/9704416].

\bibitem{linde}
A.D.~Linde,
{\it Infrared problem in thermodynamics of the Yang-Mills gas,}
Phys.\ Lett.\ {B 96} (1980) 289.

\bibitem{gpy}
D.J.~Gross, R.D.~Pisarski and L.G.~Yaffe,
{\it QCD and instantons at finite temperature,}
Rev.\ Mod.\ Phys.\ {53} (1981) 43.

\bibitem{dr}
P. Ginsparg, 
{\it First and second order phase transitions 
in gauge theories at finite temperature,}
Nucl.\ Phys.\ B 170 (1980) 388;
%
T. Appelquist and R.D. Pisarski,
{\it High-temperature Yang-Mills theories and three-dimensional 
Quantum Chromodynamics,}
Phys.\ Rev.\ D 23 (1981) 2305.

\bibitem{gsixg}
K.~Kajantie, M.~Laine, K.~Rummukainen and Y.~Schr\"oder,
{\it The pressure of hot QCD up to $g^6 \ln (1/g)$},
Phys.\ Rev.\ D, in press 
[hep-ph/0211321]. 

\bibitem{sun}
Y.~Schr\"oder, {\it Logarithmic divergence in the energy density
of the three-dimensional Yang--Mills theory}, 
in preparation. 

\bibitem{latt}
K.~Kajantie, M.~Laine, K.~Rummukainen and Y.~Schr\"oder,
{\it Measuring infrared contributions to the QCD pressure,}
Nucl.\ Phys.\ B (Proc.\ Suppl.)\  {106} (2002) 525
[hep-lat/0110122];
%
{\it Four-loop logarithms in 3D gauge + Higgs theory,}
hep-lat/0209072.

\bibitem{sd}
K.~Kajantie, M.~Laine and Y.~Schr\"oder,
{\it A simple way to generate high order vacuum graphs,}
Phys.\ Rev.\ D {65} (2002) 045008
[hep-ph/0109100].

\bibitem{ma}
M.~Achhammer, 
{\it The QCD Partition Function at High Temperatures}, 
PhD thesis, University of Regensburg, July 2000 
(Logos-Verlag, Berlin, 2001).

\bibitem{ys_radcor}
Y.~Schr\"oder, 
{\it Automatic reduction of four-loop bubbles,}
Nucl.\ Phys.\ B (Proc.\ Suppl.)\ 116 (2003) 402
[hep-ph/0211288].

\bibitem{jamv}
J.A.M.~Vermaseren,
{\it New features of FORM,}
math-ph/0010025; \newline
{\tt http://www.nikhef.nl/\~{}form/}.

\bibitem{pi}
K.G.~Chetyrkin and F.V.~Tkachov,
{\it Integration by parts: the algorithm
to calculate beta functions in 4 loops,}
Nucl.\ Phys.\ B {192} (1981) 159;
F.V.~Tkachov,
{\it A theorem on analytical calculability of
four loop renormalization group functions,}
Phys.\ Lett.\ B {100} (1981) 65.

\bibitem{laporta}
S.~Laporta,
{\it High-precision calculation of multi-loop
Feynman integrals by  difference equations,}
Int.\ J.\ Mod.\ Phys.\ A {15} (2000) 5087
[hep-ph/0102033].

\bibitem{broadhurst_qed}
D.J.~Broadhurst,
{\it Three loop on-shell charge renormalization without integration: 
$\Lambda^\rmi{$\msbar$}_\rmi{QED}$ to four loops,}
Z.\ Phys.\ C {54} (1992) 599.

\bibitem{framework}
K.~Farakos, K.~Kajantie, K.~Rummukainen and M.~Shaposhnikov,
{\it 3d physics and the electroweak phase transition: 
a framework for lattice Monte Carlo analysis,}
Nucl.\ Phys.\ {B 442} (1995) 317 [hep-lat/9412091].

\bibitem{contlatt}
M.~Laine and A.~Rajantie,
{\it Lattice-continuum relations for 3d SU(N)+Higgs theories,}
Nucl.\ Phys.\  {B 513} (1998) 471
[hep-lat/9705003].

\bibitem{mincer}
S.A.~Larin and J.A.M.~Vermaseren,
{\it The three loop QCD beta function and anomalous dimensions,}
Phys.\ Lett.\ B {303} (1993) 334
[hep-ph/9302208];
S.A.~Larin, F.V.~Tkachov and J.A.M.~Vermaseren,
{\it The FORM version of Mincer,}
preprint NIKHEF-H-91-18.

\bibitem{bn_phi4}
E.~Braaten and A.~Nieto,
{\it Effective field theory approach to high temperature thermodynamics,}
Phys.\ Rev.\ D {51} (1995) 6990
[hep-ph/9501375].

\bibitem{av}
A.~Vuorinen,
{\it Quark number susceptibilities of hot QCD up to $g^6 \ln (g)$,}
Phys.\ Rev.\ D, in press 
[hep-ph/0212283].

\bibitem{bir}
J.P.~Blaizot, E.~Iancu and A.~Rebhan,
{\it On the apparent convergence of perturbative QCD at high temperature},
hep-ph/0303045.

\bibitem{ck}
G.~Cveti\v{c} and R.~K\"ogerler,
{\it Resummations of free energy at high temperature,}
Phys.\ Rev.\  D 66 (2002) 105009
[hep-ph/0207291].

\bibitem{rvs}
M.~Strickland,
{\it Reorganizing finite temperature field theory,}
Int.\ J.\ Mod.\ Phys.\ A {16S1C} (2001) 1277;
%
E.~Braaten,
{\it Thermodynamics of hot QCD,}
Nucl.\ Phys.\  A 702 (2002) 13;
%
A.~Peshier,
{\it Resummation of the QCD thermodynamic potential,}
Nucl.\ Phys.\ A 702 (2002) 128
[hep-ph/0110342];
%
J.O.~Andersen, 
{\it Hard thermal loops and QCD thermodynamics,}
hep-ph/0210195;
%
J.P.~Blaizot, E.~Iancu and A.~Rebhan,
{\it Thermodynamics of the high-temperature quark gluon plasma},
hep-ph/0303185.


\bibitem{pert}
K.~Farakos, K.~Kajantie, K.~Rummukainen and M.E.~Shaposhnikov,
{\it 3d physics and the electroweak phase transition: perturbation theory,}
Nucl.\ Phys.\ B {425} (1994) 67
[hep-ph/9404201].

\bibitem{cs}
P.N.~Tan, B.~Tekin and Y.~Hosotani,
{\it Maxwell-Chern-Simons scalar electrodynamics at two loops,}
Nucl.\ Phys.\ B {502} (1997) 483
[hep-th/9703121].

\bibitem{joa}
J.O.~Andersen,
{\it 3d effective field theory for finite temperature scalar electrodynamics,}
Phys.\ Rev.\ D {59} (1999) 065015
[hep-ph/9709418].

\bibitem{u1big}
K.~Kajantie, M.~Karjalainen, M.~Laine and J.~Peisa,
{\it Three-dimensional U(1) gauge + Higgs theory as
an effective theory for  finite temperature phase transitions,}
Nucl.\ Phys.\ B {520} (1998) 345
[hep-lat/9711048].

\bibitem{kkb}
B.~Kastening, H.~Kleinert and B.~Van den Bossche,
{\it Three-loop ground-state energy of O(N)-symmetric
Ginzburg-Landau theory above $T_c$ in 4-$\epsilon$ dimensions
with minimal subtraction,}
Phys.\ Rev.\ B 65 (2002) 174512
[cond-mat/0109372].

\bibitem{laporta2}
S.~Laporta,
{\it High-precision $\epsilon$-expansions of massive four-loop vacuum bubbles,}
Phys.\ Lett.\ B {549} (2002) 115
[hep-ph/0210336].

\bibitem{dd}
A.~Devoto and D.W.~Duke,
{\it Table of integrals and formulae for Feynman diagram calculations,}
Riv.\ Nuovo Cim.\  {7N6} (1984) 1.

\bibitem{aleksi}
A.~Vuorinen, 
{\it Four-loop Feynman diagrams in three dimensions}, \newline
Master's Thesis, Helsinki University, 2001
(unpublished) \newline
[http://ethesis.helsinki.fi/julkaisut/mat/fysii/pg/vuorinen/fourloop.pdf].

\bibitem{gkp}
S.~Groote, J.G.~Korner and A.A.~Pivovarov,
{\it Configuration space based recurrence 
relations for sunset-type  diagrams,}
Eur.\ Phys.\ J.\ C {11} (1999) 279
[hep-ph/9903412].

\bibitem{ar_3loop}
A.K.~Rajantie,
{\it Feynman diagrams to three loops in three-dimensional field theory,}
Nucl.\ Phys.\ B {480} (1996) 729; 
{\em ibid.}\  B {513} (1996) 761 (E)
[hep-ph/9606216].

\end{thebibliography}
\end{document}